\documentclass{IEEEtran}

\usepackage{array} % 引入 array 包以使用 p{} 列格式,设置表的列宽, \tiny, \scriptsize, \footnotesize, \small, \normalsize, \large, \Large, \LARGE, \huge,
\usepackage{graphicx} % 添加图片
\usepackage{subcaption} % 添加子图
\usepackage{amsmath} % 公式对齐包
\usepackage{amsfonts} % 可以用\mathbb{E}
\usepackage{cite}
\usepackage{caption}
\usepackage{balance}
\usepackage[ruled,vlined,linesnumbered,ruled]{algorithm2e}
\usepackage{lettrine} % 第一段首字母下沉
\usepackage{hyperref} % 点击索引到达参考文献和图片
\hypersetup{hypertex=true,
	colorlinks=true,
	linkcolor=blue,
	anchorcolor=blue,
	citecolor=blue}
\begin{document}
	
\title{Transmission Scheduling of Millimeter Wave Communication for High-Speed Railway in Space-Air-Ground Integrated Network}

%\author{
%\IEEEauthorblockN{
%	Lei Liu, 
%	Yong Niu,	
%	and Bo Ai}	\\
%%	\IEEEauthorblockA{State Key Laboratory of Advanced railway Autonomous Operation, Beijing Jiaotong University, Beijing, China}
%%	\IEEEauthorblockA{E-mail: \{leiliu21, niuyong, boai\}@bjtu.edu.cn}
%%		mmWave{-0.8cm}		
%	}
\author{Lei Liu, Bo Ai, \emph{Fellow, IEEE,} Yong Niu, \emph{Senior Member, IEEE}, Zhu Han, \emph{Fellow, IEEE},\\ Ning Wang, \emph{Member, IEEE}, Lei Xiong, and Ruisi He, \emph{Senior Member, IEEE} 	

\thanks{Copyright (c) 2024 IEEE. Personal use of this material is permitted. However, permission to use this material for any other purposes must be obtained from the IEEE by sending a request to pubs-permissions@ieee.org. Manuscript received 12 June 2024; revised 27 July 2024 and 2 October 2024; accepted 14 October 2024. This work was supported in part by Beijing Natural Science Foundation under Grant L232042; in part by the National Key Research and Development Program of China under Grant 2021YFB2900301; in part by the National Natural Science Foundation of China under Grant 62221001, Grant 62231009 and Grant U21A20445; in part by the Fundamental Research Funds for the Central Universities 2023JBMC030; in part by the Fundamental Research Funds for the Central Universities under Grant 2022JBXT001 and Grant 2022JBQY004;  in part by NSF under Grant CNS-2107216, Grant CNS-2128368, Grant CMMI-2222810, and Grant ECCS-2302469, and in part by the US Department of Transportation, Toyota,	Amazon and JST ASPIRE under Grant JPMJAP2326. \emph{(Corresponding authors: Bo Ai; Yong Niu.)}
	
Lei Liu, Bo Ai, Yong Niu, Lei Xiong and Ruisi He are with the  School of Electronic and Information Engineering, Beijing Jiaotong University, Beijing 100044, China, and also with the Beijing Engineering Research Center of High-speed Railway Broadband Mobile Communications, Beijing 100044, China (e-mail: leiliu21@bjtu.edu.cn;  boai@bjtu.edu.cn;	niuy11@163.com; lxiong@bjtu.edu.cn; ruisi.he@bjtu.edu.cn).
		
Zhu Han is with the Department of Electrical and Computer Engineering, University of Houston, Houston, TX 77004 USA, and also with the Department of Computer Science and Engineering, Kyung Hee University, Seoul 446-701, South Korea (e-mail: hanzhu22@gmail.com).
	
Ning Wang is with the School of Information Engineering, Zhengzhou University, Zhengzhou 450001, China (e-mail: ienwang@zzu.edu.cn).		
		}}
\maketitle
\thispagestyle{empty} % no page number for the first page

\begin{abstract}
	
The space-air-ground integrated network (SAGIN) greatly improves coverage and reliability for millimeter-wave (mmWave) communication in high-speed railway (HSR) scenarios. However, a significant challenge arises in the transmission scheduling due to the rapid changes in channel state, link selection for train mobile relays (MRs), and order of the flow scheduling. To tackle this challenge, we introduce an optimization problem focused on maximizing the weighted sum completed flows that satisfy the quality of service (QoS) requirements for HSR mmWave communication in SAGIN. To facilitate the simultaneous scheduling of flows by base station-MR (BS-MR), satellite-airship-MR, and satellite-MR links, we propose a link selection algorithm, which can help each flow choose a suitable set of links in every frame and determine whether the BS networks need the assistance of the satellite and airship. Furthermore, taking into account the priority and occupied time slots (TSs) resource of different flows, we propose a multi-link weighted flow scheduling (MWFS) algorithm. This algorithm not only prioritizes scheduling high-priority flows but also aims to maximize the weighted sum completed flows for MRs. Our simulation results confirm that the proposed algorithm significantly increases the weighted sum completed flows and the total transmitted bits. Additionally, the proposed algorithm can achieve the optimal flow transmission in different link switching periods and enhance the scheduling of high-priority flows compared to other algorithms.		

\end{abstract}


\section{Introduction}
\lettrine[lines=2]{H}{IGH-SPEED} railway (HSR) has emerged as an important part of modern transportation systems, characterized by its high safety, rapid speeds, and large capacity to transport thousands of passengers simultaneously \cite{Ai1},\cite{Trajectory}. The HSR provides numerous benefits but also faces challenges, particularly in satisfying the high data rates, exceptional reliability, and low latency required to support its operations and passengers efficiently \cite{Gbps},\cite{Lay}. Traditionally, HSR communication systems relied on technologies such as Global System for Mobile Communications-Railway (GSM-R) and Long-Term Evolution-Railway (LTE-R) \cite{heruisi}. GSM-R, developed specifically for railway communications, supports vital features like group calls and emergency services, which are essential for safety and operations. LTE-R, an enhancement over GSM-R, provides significantly improved data speeds reaching up to hundreds of Mbps and expanded bandwidth of up to 20 MHz \cite{20MHz},\cite{zhoutao}. This upgrade provides a more robust framework for railway communications, supporting a variety of data-intensive applications. Yet, even these enhanced rates might not be sufficient for advanced applications that demand exceptionally high throughput, such as streaming high-definition video in real time or managing thousands of simultaneous connections \cite{chenwei},\cite{Challenges1}.

To meet the growing data needs in HSR, millimeter-wave (mmWave) technology has emerged as a major breakthrough. Operating within the 30 GHz to 300 GHz frequency bands, mmWave can deliver data speeds up to several gigabits per second \cite{beamforming2,AI,guanke1}. This capability is crucial for offering the strong and fast connectivity that HSR passengers require. Moreover, mmWave technology offers an extensive bandwidth, typically ranging from several hundred megahertz to several gigahertz \cite{5GmmWave},\cite{guanke2}. This significant bandwidth capacity allows mmWave to effectively meet the high data demands of HSR communications, ensuring rapid and reliable connectivity.

However, the application of mmWave technology in HSR faces several challenges, notably high path loss and significant penetration loss, which severely impact coverage of terrestrial base stations (BSs) \cite{Doppler}, \cite{BeamManagement}. To combat these issues, innovative technological solutions have been developed. Beamforming technology is employed to focus signal energy into narrower beams, thereby enhancing reach and minimizing interference, which is particularly crucial in the dynamic environment of HSR \cite{Beamforming3,Beamforming4,BeamSelection}. Additionally, the use of external relays mounted on train carriages plays a vital role in maintaining stable communication, effectively overcoming penetration of train body \cite{mayunhan2}. Moreover, the space-air-ground integrated network (SAGIN) significantly expands coverage and enhances HSR communications robustness, effectively compensating for the inherent limitations of mmWave technology such as its limited coverage range and sensitivity to obstructions \cite{SGIN-Safety},\cite{shenxuemin}.

The SAGIN contains satellite networks, aerial access networks and terrestrial networks to ensure continuous and stable communication capability of HSR, even in remote areas \cite{SAGIN2}. The aerial access networks are provided by high-altitude platforms (HAPs), such as airships and unmanned aerial vehicles (UAVs) \cite{HAP}, \cite{3gpp.38.821}. Furthermore, the SAGIN offers the flexibility to dynamically adjust resource allocation based on varying communication needs and a large number of communication links \cite{ResourceAllocationinSAGIN},\cite{RA1}. However, the implementation of SAGIN is not without its challenges, particularly in the management of prioritizing access and transmission scheduling in the mobile network \cite{satelite1}. Constant monitoring of variables such as the train's speed, real-time location of the satellite and mobile relays (MRs), and quality of service (QoS) requirements is necessary to dynamically manage time resources for SAGIN, which requires complex algorithms and robust backend processing capabilities \cite{wangyibing}. The importance of efficient transmission scheduling cannot be overstated, as it is crucial for maintaining high-quality communication services and ensuring the seamless operation of HSR systems.

In this paper, we address the transmission scheduling problem faced by HSR mmWave communication in SAGIN. The primary issue we tackle is the selection of the optimal link for MRs, with the objective of maximizing the weighted sum flows completed by adjusting the scheduling sequence of flows in all links and allocating the time slots (TSs). Then we propose a link selection algorithm and a multi-link weighted flow scheduling (MWFS) algorithm, which is engineered to complete the maximum number of flows, while also ensuring that priority transmission for high-priority flows. The contributions of this paper are summarized as follows:
\begin{itemize}
\item[$\bullet$] To address the transmission scheduling issues in SAGIN, several factors are considered, including the high-speed movement of the train and satellite, the channel variations, the QoS requirements of flows, the selection of links for MRs, the priority of flows, and the scheduling sequence for flows. Subsequently, we introduce an optimization problem aimed at maximizing the weighted sum completed flows that meet the QoS requirements for HSR mmWave communications in SAGIN.
\item[$\bullet$] We propose a link selection algorithm that can choose a set of appropriate links for MRs in each frame. Additionally, this algorithm can determine whether the assistance of the satellite and the airship is needed when the BS network serves the MRs.
\item[$\bullet$] Based on the link selection algorithm, we propose a MWFS algorithm that compares the priority of different flows and the amount of TSs resources occupied by each flow to adjust the scheduling order of the flows. This algorithm serves to effectively increase the weighted sum completed flows that satisfy the QoS requirements, while making the flows with high-priority in each frame to be scheduled first.
\item[$\bullet$] We assess the effectiveness of the proposed algorithm through simulations. The results indicate that our algorithm performs best in scheduling both the weighted sum completed flows and the total transmitted bits, compared to other algorithms in the existing literature. Furthermore, even when the link switching period is extended, our algorithm continues to outperform alternative methods in terms of overall performance.
\end{itemize}

The rest of the paper is organized as follows. In Section II, we provide an detailed overview of the related work. In Section III, we introduce the system model for the HSR mmWave communication in SAGIN. In Section IV, we formulate the problem of maximizing the weighted sum completed flows. The proposed scheduling scheme is presented in Section V and performance evaluation in Section VI. Finally, Section VI concludes this paper.

\section{RELATED WORK}

The 3GPP protocol outlines various satellite orbits and their respective roles in non-terrestrial networks, emphasizing their altitude ranges and coverage capabilities \cite{3gpp.38.811}. Low-Earth Orbit (LEO) satellites operate between 300 and 1,500 km with a circular orbit, while Medium-Earth Orbit (MEO) satellites are positioned at altitudes ranging from 7,000 to 25,000 km, also in circular orbits. Geostationary Earth Orbit (GEO) satellites are much higher at 35,786 km, maintaining a fixed position relative to a given point on earth. High Elliptical Orbit (HEO) satellites vary significantly in altitude from 400 to 50,000 km, moving in elliptical orbits around the earth. Specifically, for LEO-based non-terrestrial access networks, the maximum distance between the satellite and user equipment varies with altitude—1,932 km at 600 km altitude and 3,131 km at 1,200 km altitude, accommodating high-speed mobility such as trains moving at 500 km/h and aircraft possibly reaching 1,200 km/h \cite{3gpp.38.821}. Moreover, the maximum bandwidth in the Ka-band offers 400 MHz for both downlink and uplink, facilitating substantial data transmission capabilities \cite{400MHz}.

Current research on the resource allocation of the SAGIN has mainly focused on energy efficiency, computational resource management, bandwidth and spectrum allocation. Tang \emph{et al}. \cite{Tang} concentrated on energy-efficient task offloading to reduce energy consumption between hybrid clouds and LEO satellites. Nguyen \emph{et al}. \cite{ResourceAllocationinSAGIN} proposed a comprehensive strategy that integrated user scheduling, partial offloading control, bit allocation, and bandwidth allocation, significantly lowering the energy use of communication systems. Cao et al. \cite{DRL} proposed a user equipment (UE)-driven deep reinforcement learning (DRL) scheme for multi-user access control in non-terrestrial networks. This approach could boost long-term throughput by enabling each UE to act as a DRL agent. Innovative solutions, such as deploying airships as mobile base stations in remote areas, were also being explored to overcome communication challenges in isolated regions and improved network coverage and reliability \cite{Airship_BS}. Despite these advances, transmission scheduling remains a notably underexplored area in SAGIN, particularly in its application to HSR mmWave communications.

\begin{figure*}[h]
	\centering
	\begin{minipage}[c]{0.9\textwidth}
		\centering
		\includegraphics[width=6.5in]{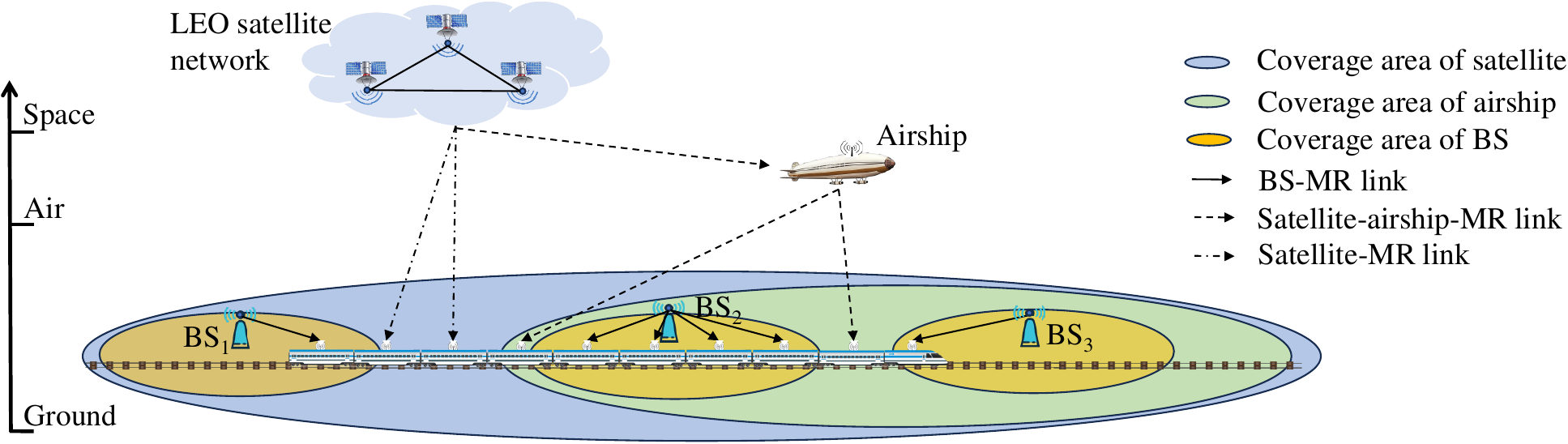}\\
		\caption{The HSR mmWave communication in SAGIN.}
		\label{HSR}
	\end{minipage} \\
\end{figure*}

Research on transmission scheduling in mmWave communications has focused on static cells and dynamic vehicular communications. Lakshmi \emph{et al}. \cite{FairScheduling} proposed the heuristic space-time division multiple access (STDMA) schedulers developed address the QoS requirements across all flows, aiming for fair concurrent transmission in mmWave wireless local area networks (WLANs). Zhu \emph{et al}. \cite{MQIS} introduced algorithms designed to maximize the number of successfully scheduled flows and improve system throughput. However, these methods do not fully tackle the mobility challenges inherent to HSR communications. In response, Wang \emph{et al}. \cite{wangyibing} developed a UAV-assisted HSR mmWave communication scheme to maximize flow number while meeting their QoS requirements. Similarly, Zhang \emph{et al}. \cite{train_ground} proposed a coalition game algorithm to optimize user association and transmission scheduling, enhancing the sum rate for HSR but not addressing flow number enhancement. Liu \emph{et al}. \cite{ll} proposed a Max-Flow Scheduling (MFS) algorithm to solve the transmission scheduling problem for an airship-assisted HSR scenario, but they do not take into account the satellite-MR link, the priority of different flows, and the influence of the satellite's position.

In summary, existing research lacks a focused study on transmission scheduling specifically for HSR mmWave communications in SAGIN. This paper introduces an optimization problem for transmission scheduling in HSR mmWave communications in SAGIN,  and proposes a link selection algorithm and a MWFS algorithm to maximize the weighted sum completed flows. 

\section{SYSTEM OVERVIEW AND ASSUMPTION}

\subsection{System Model}

We consider a downlink communication with mmWave for HSR in SAGIN that has the BS-MR, satellite-MR, satellite-airship-MR links can serve the train MRs, as shown in Fig. \ref{HSR}. The train travels at a speed $v_{t}$ on the track, sequentially passing through BS$_1$, BS$_2$, and BS$_3$. The MRs are equipped on the roof of the train with height $h^{MR}$. When the BS-MR link cannot satisfy the QoS requirements of all flows, the satellite-MR and  satellite-airship-MR links will schedule flows to the MRs to alleviate the load of the BS networks. The primary role of the airship is to act as a relay between the satellite and the MRs, focusing on signal forwarding and enhancement. In each frame, the airship reports request flows from MRs and their QoS requirements within its coverage area to the satellite, helping the satellite efficiently schedule TS resources. As a relay, the complete downlink data transmission chain for the airship is from the satellite to the airship and then to the MRs. The airship remains stationary, positioned between BS$_2$ and BS$_3$ at a height of $h^A$ and the coverage of the airship includes the coverage of BS$_2$ and BS$_3$. Moreover, from the time the train enters the coverage area of BS$_1$ until it leaves the coverage area of airship, the MRs are continuously within the coverage range of the LEO satellite network. In the BS-MR, satellite-MR, and satellite-airship-MR links, since the receivers for all three links are MRs, the penetration loss of the mmWave signal through the train body can be ignored. Additionally, because the train moves along a fixed track with a known and predictable direction and speed, the Doppler shifts between the transmitter and the MRs can be determined and mitigated using the existing technique \cite{Doppler}.

\begin{figure}[t]
	\centering
	\begin{minipage}[c]{0.5\textwidth}
		\centering
		\includegraphics[width=2.6in]{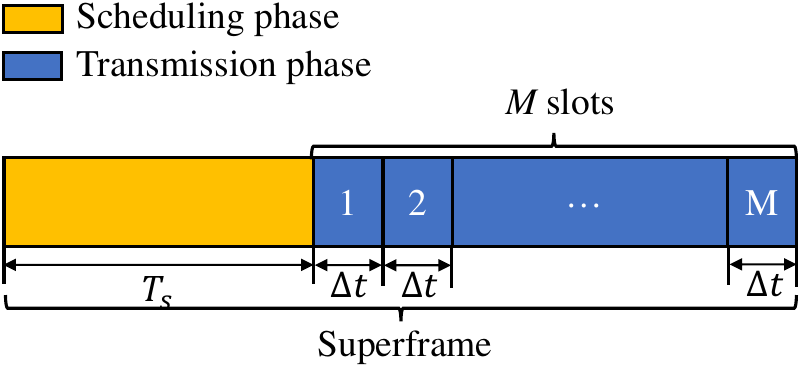}\\
		\caption{The structure of one superframe.}
		\label{frame}
	\end{minipage} \\
\end{figure}

We define $N^{frame}$ superframes as the duration in which the train travels from entering the coverage of BS$_1$ to exiting the coverage of BS$_3$. Each frame consists of a scheduling phase and a transmission phase, as illustrated in Fig. \ref{frame}. The scheduling phase is responsible for collecting requests from the train's MRs and its time length is $T_s$. In the transmission phase, the time is segmented into $M$ equal TSs. Each TS, having a duration of $\Delta t$, is allocated for data transmission. Consequently, the total duration of each frame is given by $T^{frame} = T_s + M \Delta t$. Assume that the channel conditions are considered stable within a frame if the train travels no more than $d^{frame}$. The number of TSs in this frame is $M = \lfloor(d^{frame}/v-T_s)/\Delta t \rfloor$. Therefore, increasing the train speed $v_t$ will lead to a decrease in the total number of TSs in the frame, which makes it more difficult for all links to complete the QoS requirements of flows. Table \ref{math} summarizes the main notation used in this paper.
\begin{table}[t]
	\captionsetup{labelsep=none}
	\captionsetup{font=footnotesize}
	\caption{\\NOTATION SUMMARY}
	\centering
	\begin{tabular}{p{2.8cm} p{5.2cm}}% 使用 p{} 指定列宽
		\hline
		Notation & Description \\
		\hline
		$v_t,v_s$  & Speed of the train and the satellite \\
		$h^{MR},h^{A},h^{S}$  & Height of MR, airship and satellite \\		
		$T^{frame}$, $T_s$, $\Delta t$  & Duration of the frame, the scheduling phase and the TS \\
		$h_{s_{i,n,m},r_{i,n,m}}$  & Total channel gain of flow $i$ from the transmitter $s$ to the receiver $r$ at TS $m$ in frame $n$ \\
		$h^L_{s_{i,n,m},r_{i,n,m}} $  & LoS channel gain of flow $i$ from the transmitter $s$ to the receiver $r$ at TS $m$ in frame $n$ \\ 
		$h^{NL}_{s_{i,n,m},r_{i,n,m}} $  & NLoS channel gain of flow $i$ from the transmitter $s$ to the receiver $r$ at TS $m$ in frame $n$ \\ 
		$K$  & Rician facotr \\	 
		$N^{s}$, $N^{flow}_n$, $M$ & Number of transmitters, number of flows in frame $n$, number of TSs  \\	
		$\alpha,\beta,\delta, \eta, \vartheta$ & BS$_1$, BS$_2$, BS$_3$, airship and satellite \\
		$G_t$, $G_r$ & Antenna gain of the transmitter and the receiver	\\				
		$P_{s_{i,n,m}}, P_{r_{i,n,m}}$ & Transmitted power and received power of flow $i$ at TS $m$ in frame $n$\\
		$\lambda$, $\zeta$, $\varepsilon $ & Wavelength of the signal, mmWave signal degradation due to atmospheric absorption and scintillation, efficiency of the transceiver design \\
		$\sigma$ & Path loss at reference distance\\	
		$\gamma$ & Path loss exponent for LoS case\\	
		$d_{s_{i,n,m},r_{i,n,m}}$ & Distance between the transmitter $s$ and the receiver $r$ at TS $m$ in frame $n$ \\
		$d^{BS_1}$, $d^{BS_2}$, $d^{BS_3}$, $d^{A}$ & Coverage distance of BS$_1$, BS$_2$, BS$_3$ and airship \\
		$d^{MR_i \rightarrow BS_j}_n$,$d^{MR_i \rightarrow A}_n$  &  Distance between the MR $i$ and the BS $j$ and the airship in frame $n$	\\
		$d_{n}^{M R_{i} \rightarrow S}$, $d_{n}^{S \rightarrow A}$ & Distance between the satellite and the MR $i$ and the airship in frame $n$\\
		$d_{n}^{M R_{i} \rightarrow S,h}$,$d_{n}^{S \rightarrow A,h}$  & Horizontal distance between the satellite and the MR $i$ and airship in frame $n$\\	
		$d^{MR}$ & Distance between MRs \\				
		$Q_{s_{i,n},r_{i,n}}$ & Binary variable to indicate whether the QoS requirement of flow $i$ for the receiver $r$ is satisfied by the transmitter $s$ in frame $n$ \\
		$u_{i,n}$ & Weight coefficient of flow $i$ in frame $n$  \\							
		$q_{s_{i, n},r_{i,n}}$, $\hat{q}_{s_{i, n},r_{i,n}}$ & Throughput of flow $i$ provided by the transmitter $s$ for the receiver $r$ in frame $n$, QoS requirement of flow $i$ in frame $n$ \\
		$R_{s_{i,n,m},r_{i,n,m}}$  & Transmission rate of flow $i$ provided by the transmitter $s$ for the receiver $r$ at TS $m$ in frame $n$  \\
		$ a_{s_{i,n,m},r_{i,n,m}}$   & Binary variable to indicate whether the  TS $m$ in frame $n$ is occupied by flow $i$ from transmitter $s$ to receiver $r$  \\
		$w_{s_{i,n},r_{i,n}}$    & Binary variable to indicate whether the transmitter $s$ in frame $n$ is occupied by flow $i$ of the receiver $r$ \\
		$SINR_{r_{i,n,m}}$ & SINR of flow $i$ for receiver $r$ at TS $m$ in frame $n$ \\
		$c_{n}^{M R_{i} \rightarrow S}$, $c_{n}^{S \rightarrow A}$ &  Intermediate variable\\
		$t_{max}$ & Total time for MRs passing through the $BH$\\		
		$R$, $M_e$, $G$ & Radius of the earth, mass of the earth and the gravitational constant\\
		\hline	
	\end{tabular}
	\label{math}
\end{table}

\subsection{Transmission Rate}
We model the mmWave channels of the BS-MR, satellite-MR and satellite-airship-MR links as Rician channels \cite{chenyali1},\cite{rician}. The channel gain, which includes the line-of-sight (LoS) and non-LoS (NLoS) components, is represented as
\begin{equation}
	\begin{aligned}
h_{s_{i,n,m},r_{i,n,m}}=& \sqrt{\frac{K}{1+K}} h_{s_{i,n,m},r_{i,n,m}}^{L} \\ &+\sqrt{\frac{\sigma}{1+K}\left(d_{s_{i,n,m},r_{i,n,m}}\right)^{-\gamma}} h_{s_{i,n,m},r_{i,n,m}}^{N L},
\end{aligned}
\end{equation}
where $K$ is the Rician factor. $\sigma$ is the path loss at reference distance.  $h^L_{s_{i,n,m},r_{i,n,m}}$ is the LoS channel gain of flow $i$ from transmitter $s$ to receiver $r$ at the $m$-th TS in the $n$-th frame, which can be represented by 
\begin{equation}
h_{s_{i,n,m},r_{i,n,m}}^{L}=\sqrt{\sigma\left(d_{s_{i,n,m},r_{i,n,m}}\right)^{-\gamma}} e^{-j \frac{2 \pi}{\lambda} d_{s_{i,n,m},r_{i,n,m}}},
\end{equation}
where $d_{s_{i,n,m},r_{i,n,m}}$ is the distance between the transmitter $s$ and the receiver $r$ at the $m$-th TS in the $n$-th frame. $\gamma$ is the path loss exponent in the LoS case, and $\lambda$ is the wavelength of signal.  $h^{NL}_{s_{i,n,m},r_{i,n,m}}$ is the corresponding NLoS channel gain of flow $i$ from transmitter $s$ to receiver $r$ at the $m$-th TS in the $n$-th frame, which follows $h^{NL}_{s_{i,n,m},r_{i,n,m}}\sim \mathcal{C N}(0,1)$.

Then the received power of flow $i$ at the $m$-th TS in the $n$-th frame can be expressed as 
\begin{equation}
	\begin{aligned}
P_{r_{i,n,m}}=  P_{s_{i,n,m}} G_t G_r |h_{s_{i,n,m},r_{i,n,m}}|^2   \zeta,
	\end{aligned}
	\label{Pr}
\end{equation}
where $P_{s_{i,n,m}}$ is the transmitted power of flow $i$ for transmitter $s$ at the $m$-th TS in the $n$-th frame. The antenna gain of the transmitter and receiver are $G_t$ and $G_r$. $|h_{s_{i,n,m},r_{i,n,m}}|^2$ is the channel attenuation power. $\zeta$ is the mmWave signal degradation due to atmospheric absorption and scintillation \cite{3gpp.38.811}.

The signal-to-interference plus noise ratio (SINR) for flow $i$ at the $m$-th TS in the $n$-th frame is represented as 
\begin{equation}
	\begin{aligned}
		SINR_{r_{i,n,m}}=  \frac{P_{r_{i,n,m}}} {I_{r_{i,n,m}}^b+N_0 W},
	\end{aligned}
\end{equation}
where $I_{r_{i,n,m}}^b=\sum_{b \in\{1, \ldots, N_n^{flow}\} \backslash\{i\}}P_{r_{i,n,m}}^b$ is the total interference power of flow $i$ at the $m$-th TS in the $n$-th frame. $P_{r_{i,n,m}}^b$ is the interference power of flow $i$ from flow $b$ at the $m$-th TS in the $n$-th frame. $ N_0 W$ is the noise power. $W$ is the bandwidth and $N_0$ is the noise power density.

The achievable data rate for flow $i$ at the $m$-th TS of the $n$-th frame can be determined using the Shannon's channel capacity formula as follows
\begin{equation}
	\begin{aligned}
R_{s_{i,n,m},r_{i,n,m}}=&\ \varepsilon  W \log _2\left( 1+SINR_{r_{i,n,m}} \right) \\
=&\ \varepsilon  W \log _2\left( 1+\frac{w_{s_{i,n},r_{i,n}}a_{s_{i,n,m},r_{i,n,m}}P_{r_{i,n,m}}}{a_{s_{b,n,m},r_{i,n,m}} I_{r_{i,n,m}}^b+N_0 W} \right),
	\end{aligned}
	\label{R} 
\end{equation}
where $\varepsilon \in (0,1)$ is the efficiency of the transceiver design. $w_{s_{i,n},r_{i,n}}$ is a binary variable to indicate whether the transmitter $s$ in the $n$-th frame is occupied by flow $i$. If so, $w_{s_{i,n},r_{i,n}}=1$; otherwise, $w_{s_{i,n},r_{i,n}}=0$. $a_{s_{i,n,m},r_{i,n,m}}$ is a binary variable to indicate whether the $m$-th TS of the $n$-th frame is occupied by flow $i$ for the transmitter $s$. If so, $a_{s_{i,n,m},r_{i,n,m}}=1$ and other flows cannot occupy this TS of the $n$-th frame; otherwise, $a_{s_{i,n,m},r_{i,n,m}}=0$. $a_{s_{b,n,m},r_{i,n,m}}$ is a binary variable indicating whether flow $i$ receives the interference from flow $b \in\{1, \ldots, N_n^{flow}\} \backslash\{i\}$ at the $m$-th TS in the $n$-th frame. Finally, the throughput of flow $i$ in the $n$-th frame can be calculated as 
\begin{equation}
q_{s_{i,n},r_{i,n}}=\frac{\sum_{m=1}^M{R_{s_{i,n,m},r_{i,n,m}} \Delta t}}{T_s+M \Delta t}.
\end{equation}

\subsection{Train Movement Model}
\begin{figure}[t]
	\centering
	\begin{minipage}[c]{0.5\textwidth}
		\centering
		\includegraphics[width=3.4in]{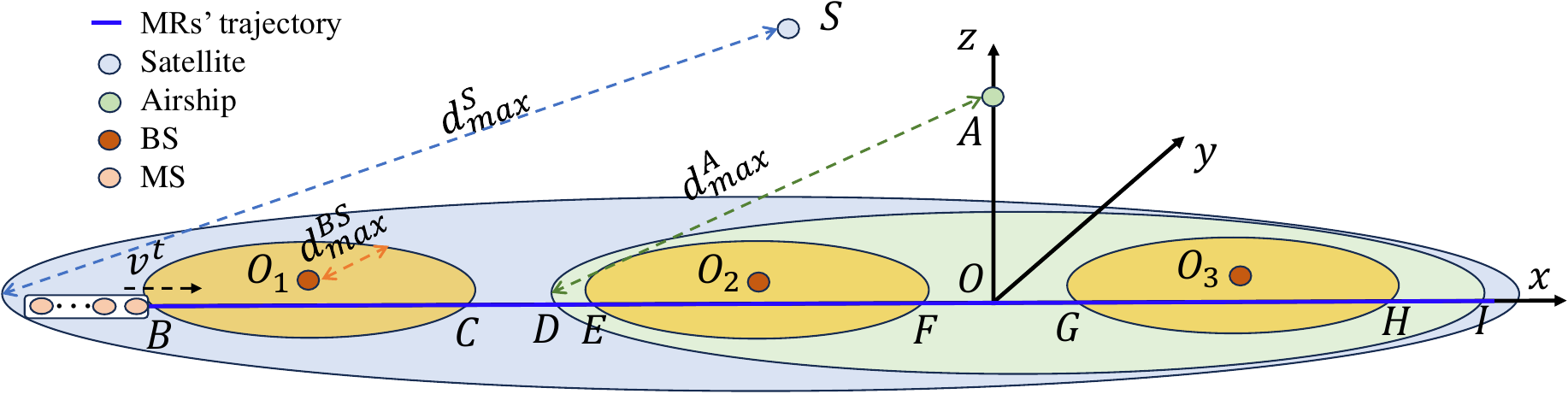}\\
		\caption{Train movement Model.}
		\label{model}
	\end{minipage} \\
\end{figure}

The trajectory of the train has a significant impact on the received signal power of the MRs. In order to facilitate the analysis of the distance between the $i$-th MR and the transmitter $s$, we approximate the trajectory of the train as a straight line, and MRs' trajectories are on the $x$-axis, as shown in Fig. \ref{model}. When the received power equals the receiver sensitivity $R_s$, the maximum coverage distances for the BS, airship, and satellite are denoted as $d_{max}^{BS}$, $d_{max}^{A}$ and $d_{max}^{S}$, respectively. Furthermore, the airship remains stationary above the coordinate origin, positioned at coordinates $(0,0,h^A)$. And the coordinates of the $i$-th MR at the $m$-th TS in the $n$-th frame are $\left( x_{n,m}^{MR_i},0,h^{MR} \right)$. The coordinates of the BS$_1$, BS$_2$ and BS$_3$ are $O_1\left(x_1,y_1,z_1 \right)$, $O_2\left(x_2,y_2,z_2 \right)$, and $O_3\left(x_3,y_3,z_3 \right)$, respectively. Besides, we set that the coordinates of the MRs entering the coverage of the BS$_1$, BS$_2$, BS$_3$ and airship are points $B$, $E$, $G$ and $D$, respectively. The coordinates of the MRs leaving the coverage of the BS$_1$, BS$_2$, BS$_3$ and airship are points $C$, $F$, $H$ and $I$, respectively. Since points $B$ and $C$ are intersections between circle $O_1$ and the $x$ axis, we have
\begin{equation}
	\begin{aligned}
\left\{ \begin{array}{l}
	\left( x-x_1 \right) ^2+\left( y-y_1 \right) ^2=\left( d_{max}^{BS} \right) ^2-\left( h^{BS}-h^{MR} \right) ^2,\\
	y=0,\\
\end{array} \right. 
	\end{aligned}
\end{equation} 
where $\left( d_{max}^{BS} \right) ^2-\left( h^{BS} - h^{MR} \right) ^2$ is the square of horizontal distance between points $O_1$ and $B$. Then the coordinates of points $B$ and $C$ can be calculated as $(x_1-\sqrt{(d_{max}^{BS})^2-(  h^{BS} - h^{MR})^2-y_1^2},0,h^{MR})$ and $(x_1+\sqrt{(d_{max}^{BS})^2-(  h^{BS} - h^{MR})^2-y_1^2},0,h^{MR})$, respectively. Repeating the above steps yields the coordinates of points $E(x_2-\sqrt{(d_{max}^{BS})^2-( h^{BS} - h^{MR})^2-y_2^2},0,h^{MR})$,  $F(x_2+\sqrt{(d_{max}^{BS})^2-(  h^{BS} - h^{MR})^2-y_2^2},0,h^{MR})$, $G(x_3-\sqrt{(d_{max}^{BS})^2-( h^{BS} - h^{MR})^2-y_3^2},0,h^{MR})$ and $H(x_3+\sqrt{(d_{max}^{BS})^2-(h^{BS} - h^{MR})^2-y_3^2},0,h^{MR})$. Then the coverage area of BS$_1$, BS$_2$, BS$_3$ and airship can be obtained as
\begin{equation}
	d^{BS_1} = 2\sqrt{(d_{max}^{BS})^2-( h^{BS}- h^{MR})^2-y_1^2},
\end{equation} 
\begin{equation}
	d^{BS_2} = 2\sqrt{(d_{max}^{BS})^2-( h^{BS}-h^{MR})^2-y_2^2},
\end{equation} 
\begin{equation}
	d^{BS_3} = 2\sqrt{(d_{max}^{BS})^2-( h^{BS}-h^{MR})^2-y_3^2},
\end{equation} 
\begin{equation}
	d^{A} = 2\sqrt{(d_{max}^{A})^2-( h_{A}-h^{MR})^2}.
\end{equation} 

Then we set the coordinates of the $i$-th MR to the $(x_1-\sqrt{(d_{max}^{BS})^2-(  h^{BS} - h^{MR})^2-y_1^2} - (i-1)d^{MR} + (n-1)d^{frame},0,h^{MR})$ in the $n$-th frame. The distance between the $i$-th MR and BS$_1$, BS$_2$, BS$_3$ and airship in the $n$-th frame can be represented as
\begin{equation}
	\begin{aligned}
		d^{MR_i \rightarrow BS_1}_n = & \Bigg( \Big(\big((d_{max}^{BS})^2 - (h^{BS} - h^{MR})^2 - y_1^2 \big)^{\frac{1}{2}} \\
	& + (i-1)d^{MR} - (n-1)d^{frame} \Big)^2 + y_1^2 \\
	& + (h^{BS} - h^{MR})^2  \Bigg)^{\frac{1}{2}},
	\end{aligned}
\end{equation} 

\begin{equation}
	\begin{aligned}
		d^{MR_i \rightarrow BS_2}_n = &  \Bigg( \Big(x_1- \big((d_{max}^{BS})^2 - (h^{BS} - h^{MR})^2 - y_1^2 \big)^{\frac{1}{2}} \\
		& - (i-1)d^{MR} + (n-1)d^{frame} - x_2 \Big)^2 + y_2^2 \\
		& + (h^{BS} - h^{MR})^2  \Bigg)^{\frac{1}{2}},
	\end{aligned}
\end{equation} 
\begin{equation}
	\begin{aligned}
		d^{MR_i \rightarrow BS_3}_n =  &  \Bigg(  \Big(x_1- \big((d_{max}^{BS})^2 - (h^{BS} - h^{MR})^2 - y_1^2 \big)^{\frac{1}{2}} \\
		& - (i-1)d^{MR} + (n-1)d^{frame} - x_3 \Big)^2 + y_3^2 \\
		& + (h^{BS} - h^{MR})^2  \Bigg)^{\frac{1}{2}},
	\end{aligned}
\end{equation}
\begin{equation}
	\begin{aligned}
		d^{MR_i \rightarrow A}_n = &  \Bigg( \Big(x_1- \big((d_{max}^{BS})^2 - (h^{BS} - h^{MR})^2 - y_1^2  \big)^{\frac{1}{2}} \\
		& - (i-1)d^{MR} + (n-1)d^{frame} \Big)^2 \\
		& + (h_{A} - h^{MR})^2  \Bigg)^{\frac{1}{2}},
	\end{aligned}
\end{equation}
respectively. The total distance for MRs passing through the coverage area of BS$_1$, BS$_2$ and BS$_3$ can be represented as
\begin{equation}
	\begin{aligned}
	B H= & x_{3}+\sqrt{\left(d_{max }^{B S}\right)^{2}-\left(h^{B S}-h^{M R}\right)^{2}-y_{3}^{2}}-x_{1} \\
	& +\sqrt{\left(d_{max }^{B S}\right)^{2}-\left(h^{B S}-h^{M R}\right)^{2}-y_{1}^{2}}.
	\end{aligned}
\end{equation}	

The total communication time for MRs passing through the $BH$ can be represented as
\begin{equation}
t_{max }=\frac{BH+\left(N^{M R}-1\right) d^{M R}}{v_t},
\end{equation}	
where $d^{MR}$ is the distance between the MRs. $N^{MR}$ is the number of the MRs. Then the number of frames during the communication time $t_{max}$ can be expressed as
\begin{equation}
N^{frame}= \left\lfloor \frac{t_{max}}
{\left(T_{s}+M  \Delta t\right) v_t}\right\rfloor.
\end{equation}

\subsection{Satellite Model}
We consider using the LEO satellite to assist in HSR mmWave communication, which can offer lower latency, improved coverage, reduced signal attenuation and enhanced mobility support. When the LEO satellite's altitude is $h^S$, the satellite's velocity can be expressed as:
\begin{equation}
v_{s}=\sqrt{\frac{G M_{e}}{R+h^{S}}},
\label{v_s}
\end{equation}	
where $G=6.674 \times 10^{-11} \mathrm{~m}^{3} \mathrm{~kg}^{-1} \mathrm{~s}^{-2}$ is the gravitational constant. $M_e$ is the mass of the earth. $R$ is the radius of the earth. According to the Kepler's Third Law \cite{Kepler}, the period for a LEO satellite with altitude $h^S$ to complete one orbit around the earth is
\begin{equation}
T^{S}=2 \pi \sqrt{\frac{\left(R+h^{S}\right)^{3}}{G M_{e}}}.
\label{T^S}
\end{equation} 

As an example, a typical altitude of the LEO satellite is $600$ km, and according to (\ref{v_s}) and (\ref{T^S}), we can obtain that the velocity of the LEO satellite is $7.5$ km/s, and the period is $96$ min/orbit \cite{3gpp.38.821}. This indicates that when the LEO satellite is used to assist HSR mmWave communications, the satellite's position changes rapidly in real time. Therefore, the impact of satellite position changes must be considered when scheduling transmission flows between the LEO and MRs.

Assuming that the position of the satellite is $S\left(l_{n}^{a, S}, l_{n}^{o, S}, h^{S}\right)$ in the $n$-th frame, where $l_{n}^{a, S}$ and $l_{n, m}^{o, S}$ represent the latitude and longitude of the satellite, respectively. To calculate the distance between the satellite and the $i$-th MR, we assume the latitude and longitude of the $i$-th MR to be $l_{n}^{a, MR_{i}}$ and $l_{n}^{o, MR_{i}}$. According to the Haversine formula \cite{Haversine}, the horizontal distance between the satellite and the $i$-th MR can be expressed as

\begin{equation}
d_{n}^{M R_{i} \rightarrow S,h}= 2 R \arcsin \left(\sqrt{c_{n}^{M R_{i} \rightarrow S}}\right),
\end{equation}	
where the intermediate variable $c_{n}^{M R_{i} \rightarrow S}$ is
\begin{equation}
	\begin{aligned}
		c_{n}^{M R_{i} \rightarrow S}= & \sin ^{2}\left(\frac{\Delta l_{n}^{a, M R_{i} \rightarrow S}}{2}\right)+\cos \left(l_{n}^{a, S}\right) \\
		& \cdot \cos \left(l_{n}^{a, M R_{i}}\right)  \sin ^{2}\left(\frac{\Delta l_{n}^{o, M R_{i} \rightarrow S}}{2}\right),
	\end{aligned}
\end{equation}	
where the differences in latitude and longitude between the $i$-th MR and satellite in the $n$-th frame are $\Delta l_{n}^{a, M R_{i} \rightarrow S}=l_{n}^{a, S}-l_{n}^{a, M R_{i}}$ and $\Delta l_{n}^{o, M R_{i} \rightarrow S}=l_{n}^{o, S}-l_{n}^{o, M R_{i}}$, respectively. Since the altitude of the satellite significantly affects the actual distance between it and the $i$-th MR, the actual distance between them in the $n$-th frame can be expressed as
\begin{equation}
d_{n}^{M R_{i} \rightarrow S}=\sqrt{\left(d_{n}^{M R_{i} \rightarrow S,h}\right)^{2}+\left(h^{S}-h^{M R_{i}}\right)^{2}}.
\end{equation}

Similarly, the horizontal distance between the satellite and the airship can be expressed as
\begin{equation}
	d_{n}^{S \rightarrow A,h}= 2 R \arcsin \left(\sqrt{c_{n}^{S \rightarrow A}}\right),
\end{equation}	
where the intermediate variable $c_{n}^{S \rightarrow A}$ is
\begin{equation}
	\begin{aligned}
		c_{n}^{S \rightarrow A}= & \sin ^{2}\left(\frac{\Delta l_{n}^{a, S \rightarrow A}}{2}\right)+\cos \left(l_{n}^{a, S}\right)\\
		& \cdot \cos \left(l_{n}^{a, A}\right) \cdot \sin ^{2}\left(\frac{\Delta l_{n}^{o, S \rightarrow A}}{2}\right),
	\end{aligned}
\end{equation}	
where the differences in latitude and longitude between the $i$-th MR and airship in the $n$-th frame are $\Delta l_{n}^{a, S \rightarrow A}=l_{n}^{a, S}-l_{n}^{a, A}$ and $\Delta l_{n}^{o, S \rightarrow A}=l_{n}^{o, S}-l_{n}^{o, A}$, respectively. The actual distance between the satellite and the airship in the $n$-th frame can be expressed as
\begin{equation}
	d_{n}^{S \rightarrow A}=\sqrt{\left(d_{n}^{S \rightarrow A,h}\right)^{2}+\left(h^{S}-h^{A}\right)^{2}}.
\end{equation}

It is important to note that if the satellite is not on the same side of the Earth as the airship or MR, satellite communication will be interrupted. Consequently, this paper only considers distance calculations for the satellite, airship, and MR when they are on the same side of the Earth.

\subsection{Flow Priority}
Considering that different flows carry various services, we can adjust the transmission scheduling strategy based on the priority and QoS requirements of each flow. Typically, services with higher priority are more sensitive to delays and require higher reliability, such as autonomous train operation and real-time train monitoring. Services with lower priority can tolerate higher delays and less stringent transmission requirements, such as web browsing and email transmission. We define the weight coefficient of the $i$-th flow as $u_{i,n}$ in the $n$-th frame, with its range of values being [0.2, 0.5, 0.8, 1]. Under the meeting of the QoS requirements, each link must prioritize to schedule the flows with a higher value of $u_{i,n}$.

\section{PROBLEM FORMULATION AND ANALYSIS}
We consider the transmission schedule problem for the HSR mmWave communication in SAGIN. Our goal is that the more weighted flows can be scheduled that satisfies the QoS requirements. If a flow’s QoS requirement is satisfied, this flow is called the completed flow. Then we define a binary variable $Q_{s_{i,n},r_{i,n}}$ to indicate whether the QoS requirement of flow $i$ for the receiver $r$ is satisfied by the transmitter $s$ in the $n$-th frame. If so, $Q_{s_{i,n},r_{i,n}}=1$; otherwise, $Q_{s_{i,n},r_{i,n}}=0$. Therefore, the optimal scheduling problem is formulated as 
\begin{equation} 
\max \sum_{s=1}^{N^{s}} \sum_{n=1}^{N^{frame}} \sum_{i=1}^{N^{flow}_n} u_{i,n} Q_{s_{i,n},r_{i,n}},
\end{equation}
where $N^{s}$ is the number of transmitters. $N^{flow}_n$ is the number of flows in the $n$-th frame.

For the BS-MR link and satellite-MR link, the condition of flow $i$ to be successfully scheduled is that the throughput provided by the transmitter $s$ is greater than the QoS requirement of flow $i$. It can be represented as 
\begin{equation}  
	\begin{aligned}
 q_{s_{i, n},r_{i,n}}= & \frac{\sum_{m=1}^M{R_{s_{i,n,m},r_{i,n,m}} \Delta t}}{T_s+M \Delta t} \geq \hat{q}_{s_{i, n},r_{i,n}}, \\ 
 & \forall i, n, s=\{\alpha, \beta, \delta, \vartheta \}, r=\{1,...,MR_{N^{MR}} \},
 	\end{aligned}
 	\label{q}
\end{equation}   	
where $s=\alpha, \beta, \delta$ and $\vartheta$ indicate that the transmitter is the BS$_1$, BS$_2$, BS$_3$ and satellite, respectively.  $\hat{q}_{s_{i,n},r_{i,n}}$ is the QoS requirement of flow $i$ in the $n$-th frame. It is important to note that the variable we need to optimize is $ a_{s_{i,n,m},r_{i,n,m}}$. The transmission rate $R_{s_{i,n,m},r_{i,n,m}}$ is influenced by the binary variable $ a_{s_{i,n,m},r_{i,n,m}}$ in (\ref{q}), as seen from (\ref{R}). Moreover, the binary variable $ a_{s_{i,n,m},r_{i,n,m}}$  determines the value of $Q_{s_{i,n},r_{i,n}}$. Besides, since each flow has its own QoS requirement, the transmitter $s$ needs to take up enough TSs to successfully schedule the flow $i$. However, the number of available TSs in each frame is at most $M$, and multiple flows may compete for $M$ TSs in a frame, so the key to scheduling more flows is to optimize the variable $ a_{s_{i,n,m},r_{i,n,m}}$.

For the satellite-airship-MR link, flow $i$ being successfully scheduled requires that the following two conditions are simultaneously satisfied
\begin{equation}  
	\begin{aligned}
q_{s _{i,n}, r_{i,n}}= & \frac{\sum_{m=1}^{M}{R_{s _{i,n,m},r_{i,n,m}}}\Delta t}{T_s+M\Delta t}\geq \hat{q}_{s _{i,n},r_{i,n}} , \\
& \forall i,n,  s = \vartheta, r=\eta,
	\end{aligned}
\end{equation} 
\begin{equation}  
	\begin{aligned}
q_{s_{i,n},r_{i,n}}= &\frac{\sum_{m=1}^{M}{R_{s _{i,n,m},r_{i,n,m}}}\Delta t}{T_s+M\Delta t}\geq \hat{q}_{s _{i,n},r_{i,n}},\\
& \forall i,n,s = \eta , r=\left\{ 1,...,MR_{N^{MR}} \right\},
	\end{aligned}
\end{equation} 
where $r=\eta$ indicates that the receiver is the airship. 

For the BS-MR link and satellite-MR link, flow $i$ is scheduled by at most one transmitter in one frame, then the binary variable $w_{s_{i,n},r_{i,n}}$ needs to satisfy
\begin{equation} 
	\begin{aligned}
	w_{\alpha_{i,n},r_{i,n}}+& w_{\beta _{i,n},r_{i,n}}+w_{\delta _{i,n},r_{i,n}}+w_{\vartheta_{i,n},r_{i,n}} \leq 1, \\
&  \forall i,n, r=\left\{ 1,...,MR_{N^{MR}} \right\}.
	\end{aligned}
\end{equation}

For the satellite-airship-MR link, flow $i$ is scheduled by at most two transmitters in one frame, then the binary variable $w_{s_{i,n},r_{i,n}}$ needs to satisfy
\begin{equation} 
	\begin{aligned}
		&w_{\vartheta _{i,n},\eta _{i,n}}+w_{\eta _{i,n},r_{i,n}} \leq 2, \\
		&  \forall i,n, r=\left\{ 1,...,MR_{N^{MR}} \right\}.
	\end{aligned}
\end{equation}

Moreover, the effective allocation of TSs to the appropriate flows within each frame is crucial for maximizing the number of schedulable flows. Based on equations (\ref{R}), the ability of the transmitter to satisfy the QoS requirements for flow $i$ is primarily influenced by the binary variable $a_{s_{i,n,m}}$. This variable determines whether the $m$-th TS of the $n$-th frame is assigned to flow $i$. Hence, optimizing this binary variable is essential for addressing optimal scheduling problem. Additionally, the binary variable $a_{s_{i,n,m}}$ must satisfy
\begin{equation} 
\sum_{i=1}^{N^{flow}_n}a_{s_{i,n,m},r_{i,n,m}} \leq 1,\ \forall s,r,n,m.
\label{a}
\end{equation}

The number of flows in each frame is satisfied as
\begin{equation} 
	0 \leq N^{flow}_n \leq N^{MR}.
	\label{Nflow}
\end{equation}

Finally, the transmission scheduling problem of maximizing the weighted sum completed flows for the HSR mmWave communication in SAGIN can be represented as
\begin{equation} 
\begin{aligned}
&	\,\,\left( \mathbf{P1} \right) \,\,\max \sum_{s=1}^{N^{s}} \sum_{n=1}^{N^{frame}}\sum_{i=1}^{N^{flow}_n} u_{i,n} Q_{s_{i,n},r_{i,n}}\\
& \text{s.t.\,\, Constraints}\left( \ref{q} \right) -\left( \ref{Nflow} \right).\\
\end{aligned}
\end{equation}

The problem $\mathbf{P1}$, a mixed-integer nonlinear programming (MINLP) problem, involves the binary variable $Q_{s_{i,n},r_{i,n}}$. Its objective function is nonlinear, incorporating a logarithmic function to model the achievable data rate of each flow \cite{wangyibing},\cite{STDMA}. The transmission schedule scheme is influenced by various factors including the MRs' movement trajectories, the QoS requirements of flows, and so on. Due to the complexity of direct solutions, we propose employing a link selection algorithm and a MWFS algorithm to address $\mathbf{P1}$. 

\section{PROPOSED SCHEDULING SCHEME}
In addressing the problem $\mathbf{P1}$, we focus on optimizing two main areas: the selection of links for MRs and the multi-link weighted flow scheduling. Then we develop two algorithms. The first is the link selection algorithm that can choose the optimal links for MRs in each frame. The second is the MWFS algorithm that can maximize the weighted sum completed flows for the HSR mmWave communication in SAGIN.

\subsection{Link Selection for Mobile Relay}
Appropriate link selection can significantly increase the number of completed flows for HSR mmWave communication in SAGIN. The choice of link is influenced by the relative positions of the MRs to the transmitter. As the distance between the MRs and the transmitter decreases, the received power and the maximum transmission rate of the MRs increase, which can contribute to meeting the QoS requirements of flows. We define that all flows of the $n$-th frame are in the set $S^{all}_n$. Since there are three links available to serve the MRs for HSR mmWave communications, we divide the link selection algorithm into the following three cases.

\begin{enumerate}
	\item   \emph{BS-MR link}: The MRs enter the coverage area of BSs, and BSs can provide throughput greater than the QoS requirement for any one flow in the $n$-th frame. This indicates that the BS-MR link is selectable. If the total throughput provided by the BSs is greater than the sum of the QoS requirements of all flows, the number of completed flows will be $N^{MR}$ by the BS-MR link and the other links will not be selected. 
	\item \emph{Satellite-airship-MR link}: The MRs are in the coverage area of the airship. Both the satellite-airship link and airship-MR link can provide throughput greater than the QoS requirement for any one flow in the $n$-th frame. This indicates that the satellite-airship-MR link can be used to serve the MRs.
	\item   \emph{Satellite-MR link}: Not all MRs are in the coverage area of the airship and BSs, or there exist some flows with QoS requirements that cannot be completed over the BS-MR link and satellite-airship-MR link. This indicates that the satellite-MR link need to be used.
\end{enumerate}

Furthermore,  based on the QoS requirements of flows, we can calculate the number of TS occupied by flow $i$ in the $n$-th frame can be expressed as
\begin{equation} 
	\varpi _{s_{i,n},r_{i,n}}=\frac{\hat{q}_{s_{i,n},r_{i,n}} \left( T_s+M \Delta t \right)}{q_{s_{i,n},r_{i,n}} \Delta t},
	\label{w}
\end{equation}
where $\hat{q}_{s_{i,n},r_{i,n}} ( T_s+M \Delta t)$ is the number of bits required for flow $i$ to be scheduled successfully by transmitter $s$ in the $n$-th frame. $q_{s_{i,n},r_{i,n}} \Delta t$ is the transmitted bits of transmitter $s$ for flow $i$ in a TS. 

\vspace{0.5pt}
The link selection algorithm is summarized in Algorithm 1, where $S_{n}^{BS}$, $S_{n}^{Satellite}$, $S_{n}^{Airship}$ denote the sets of flows that are in the coverage of the BSs, satellite and airship in the $n$-th frame, respectively. With the movement of the train and satellite, the distances between the MRs and BSs, airship and satellite are different in the different frames. When the distance between the $i$-th MR and BS is less than the coverage area of the BS in the $n$-th frame, it means that the $i$-th MR enters the coverage area of BS and its flow will be recorded in the set $S_{n}^{BS}$. Meanwhile, the number of TS occupied by flow $i$ is $\varGamma_{\beta_{n}}, \varGamma_{\eta_{n}} , \varGamma_{\delta_{n}}$ for BS$_1$, BS$_2$ and BS$_3$ in the $n$-th frame. Similarly, the flows that can be scheduled in the coverage of the airship and satellite are recorded in the set $S_{n}^{Airship}$ and $S_{n}^{Satellite}$. When there is any BS that can satisfy the QoS requirements of all flows in the $n$-th frame, we set $\widehat{S}_{n}^{BS} = S_{n}^{BS}, \widehat{S}_{n}^{Airship} = \emptyset$ and $\widehat{S}_{n}^{Satellite} = \emptyset$. Otherwise, the satellite-airship-MR link and the satellite-MR link are required to schedule the remain flows for MRs. Besides, to enhance the weighted sum completed flows, we further propose a MWFS algorithm in the next section.

\begin{algorithm}[!t]	
	\caption{Link Selection Algorithm}
	\label{Algorithm1}
	\SetKwData{Or}{\textbf{or}}
	\DontPrintSemicolon
	%\SetAlgoLined
	\KwIn {$S_{n}^{BS}=S_{n}^{Satellite}=S_{n}^{Airship}=\emptyset$, $\varpi _{s_{i,n},r_{i,n}}$, $\varGamma_{\alpha_{n}} =\varGamma_{\beta_{n}} = \varGamma_{\delta_{n}}=0$;}
	\KwOut {$S_{n}^{BS}, {S}_{n}^{Satellite}, {S}_{n}^{Airship}$;}
	\For{frame $n (1 \leq n <  N^{frmae})$}
	{
		\For{flow  $i (1 \leq i \leq |S^{all}_n|)$}{
			Calculate the distances $d^{MR_i \rightarrow BS_1}_n$, $d^{MR_i \rightarrow BS_2}_n$, $d^{MR_i \rightarrow BS_3}_n$, $d^{MR_i \rightarrow A}_n$, $d^{S \rightarrow A}_n$, and $d^{MR_i \rightarrow S}_n$;\\
			\If{$\min \{d^{MR_i \rightarrow BS_1}_n, d^{MR_i \rightarrow BS_2}_n, d^{MR_i \rightarrow BS_3}_n \} \leq d_{max}^{BS}$}
			{
				$S_{n}^{BS} = S_{n}^{BS} \cup flow_i$; \\
				$\varGamma_{\alpha_{n}} = \varGamma_{\alpha_{n}} + \varpi_{\alpha_{i,n},r_{i,n}}$;\\
				$\varGamma_{\beta_{n}} = \varGamma_{\beta_{n}} + \varpi_{\beta_{i,n},r_{i,n}}$;\\
				$\varGamma_{\delta_{n}} = \varGamma_{\delta_{n}} + \varpi_{\delta_{i,n},r_{i,n}}$;
			}
			\If{$ d^{MR_i \rightarrow A}_n \leq d_{max}^{A}$, $ d^{S \rightarrow A}_n \leq d_{max}^{S}$}
			{
				$S_{n}^{Airship} = S_{n}^{Airship} \cup flow_i$; \\
			}
			\If{$ d^{MR_i \rightarrow S}_n \leq d_{max}^{S}$}
			{
				$S_{n}^{Satellite} = S_{n}^{Satellite} \cup flow_i$;
			}
		
		}
		\If{$\min \{\varGamma_{\alpha_{n}}, \varGamma_{\beta_{n}}, \varGamma_{\delta_{n}} \} \leq M$}
		{
			${S}_{n}^{Airship} = \emptyset$; \\	
			${S}_{n}^{Satellite} = \emptyset$; 
		}
	}	
\end{algorithm}

\subsection{Multi-link Weighted Flow Scheduling Algorithm}
When the BS-MR link cannot meet the QoS requirements of all flows, the satellite-airship-MR link and the satellite-MR link can assist the BS-MR link in scheduling more flows for MRs. However, if each link schedules flows for the MR randomly, some flows with higher QoS requirements may be prioritized for scheduling, resulting in a decrease in the weighted flows scheduled. To increase the total number of flows scheduled when the network is congested, we propose a MWFS algorithm. This algorithm reorders the flow scheduling sequence among the three links while considering the priority of flows. 

Considering that the priority of different flows in the $n$-th frame may be different, we require that high priority flows are scheduled first. Since there are flows prioritized into five categories, we sort the flows in the sets $S_{n}^{BS}$, ${S}_{n}^{Satellite}$ and ${S}_{n}^{Airship}$ by priority of flows and sort these flows with the same priority by the number of occupied TSs. Then we record them into the set $\widehat{S}_{n}^{BS_j}$, $\widehat{S}_{n}^{Airship}$ and $\widehat{S}_{n}^{Satellite}$. $\widehat{S}_{n}^{BS_j}$ denotes the order of the $j$-th BS scheduling flow $i$ with priority $u$ in the $n$-th frame, and the number of TSs occupied by flows in the set $\widehat{S}_{n}^{BS_j}$ is recorded in the set $E_{n}^{BS_j}$. $\widehat{S}_{n}^{Airship}$ denotes the order of the airship scheduling flow $i$ with priority $u$ in the $n$-th frame, and the number of TSs occupied by flows in the set $\widehat{S}_{n}^{Airship}$ is recorded in the set $E_{n}^{Airship}$. $\widehat{S}_{n}^{Satellite}$ denotes the order of the satellite scheduling flow $i$ with priority $u$ in the $n$-th frame, and the number of TSs occupied by flows in the set $\widehat{S}_{n}^{Satellite}$ is recorded in the set $E_{n}^{Satellite}$. Besides, with the movement of the train, we set the period of link switching to $T_p = KT^{frame}$, where $K \geq 1$ is the integer.

The MWFS algorithm is shown in the Algorithm \ref{2-1}, which is categorized into link switching state and non-link switching state. The link switching period is determined by variable $K$. For example, variable $K=2$ mean that the link is switched every two frames. If frame $n=1+oK$, the network will be in the switching state and each MR will reselect the link to receive the flow. We denote that the BS$_1$, BS$_2$, BS$_3$, airship and satellite can use the same number of initial TS, then $r_{n}^{BS_1}=r_{n}^{BS_2}=r_{n}^{BS_3}=r_{n}^{A}=r_{n}^{S}=M$. For the BS-MR link, if the set $\widehat{S}_{n}^{BS_j} \neq \emptyset$, the $j$-th BS can successfully schedule at least one flow for MRs. Then we let the $j$-th BS to schedule the flow $i$ in the set $\widehat{S}_{n}^{BS_j}$ sequentially until $r_{n}^{BS_j}- E_{n}^{BS_j} < 0$ or all the flows are scheduled successfully. For the satellite-airship-MR link, it mainly schedules flows that are not successfully scheduled by BS-MR link. If the set $\widehat{S}_{n}^{Airship} \neq \emptyset $, it means that satellite-airship-MR link can schedule the flows to the MRs, and these flows are set to $\widehat{S}_{n}^{Airship} = \widehat{S}_{n}^{Airship} - \widetilde{S}_{n}^{BS_j}$. Then we schedule the flows in the set $\widehat{S}_{n}^{Airship}$ sequentially and determine whether the flows can be scheduled successfully. If the remaining TS is $r_{n}^{S} - E_{n}^{Satellite}\geq0$, it means that the satellite-airship link can complete the QoS requirement of flow $i$ in the $n$-th frame; if the remaining TS is $r_{n}^{A} - E_{n}^{Satellite}-E_{n}^{Airship}\geq0$, the airship-MR link can complete the QoS requirement of flow $i$ in the $n$-th frame. Only if both satellite-airship link and airship-MR link are schedulable for flow $i$, satellite-airship-MR link can schedule the flow $i$ successfully in the $n$-th frame. Otherwise, the satellite-MR link will schedule the remaining flows to the MRs. The flows that are successfully scheduled in the above process for the BS-MR link, satellite-airship-MR link and satellite-MR link are recorded in the sets $\widetilde{S}_{n}^{BS_j}$, $\widetilde{S}_{n}^{Airship}$ and $\widetilde{S}_{n}^{Satellite}$, respectively. If the flow $i$ in any link is scheduled successfully, the weighted sum completed flows are set to $u_{i,n}Q_{s_{i,n},r_{i,n}}=u_{i,n}$.

\begin{algorithm}[!t]	
	\renewcommand{\thealgocf}{2-1}
	\caption{MWFS Algorithm}
	\label{2-1}
	\SetKwData{Or}{\textbf{or}}
	\DontPrintSemicolon
	%\SetAlgoLined
	\KwIn {$\widehat{S}_{n}^{BS_j}$, $\widehat{S}_{n}^{Airship}$, $\widehat{S}_{n}^{Satellite}$, $u_{i,n}Q_{s_{i,n},r_{i,n}}=0$, $\widetilde{S}_{n}^{BS_j}=\widetilde{S}_{n}^{Airship}=\widetilde{S}_{n}^{Satellite} = \emptyset$, $o=0$, $K$;}
	\KwOut {$u_{i,n}Q_{s_{i,n},r_{i,n}}$;}
	\For{frame $n (1 \leq n \leq N^{frame})$}
	{
		$r_{n}^{BS_1}=r_{n}^{BS_2}=r_{n}^{BS_3}=r_{n}^{A}=r_{n}^{S}=M$;\\
		\If{$n=1+oK$}
		{
			\For {BS j($1 \leq j \leq 3$) }
			{
				\If{ $\widehat{S}_{n}^{BS_j} \neq \emptyset $}
				{					

					\For{ flow $i(1 \leq i \leq |\widehat{S}_{n}^{BS_j}|)$ }
					{
						$r_{n}^{BS_j} = r_{n}^{BS_j} - E_{n}^{BS_j}$; \\
						\If{$r_{n}^{BS_j} \geq 0$ }
						{
							$u_{i,n}Q_{s_{i,n},r_{i,n}}=u_{i,n}$;\\
							$\widetilde{S}_{n}^{BS_j} = \widetilde{S}_{n}^{BS_j} \cup \widehat{S}_{n}^{BS_j}(i)$;
						}
						\Else
						{
							Inverse process for line $7$;							
						}
					}
				}				
			}	
			$\widetilde{S}_{n}^{BS}= \widetilde{S}_{n}^{BS_1}+\widetilde{S}_{n}^{BS_2}+\widetilde{S}_{n}^{BS_3}$;\\
			$\widehat{S}_{n}^{Airship} =  \widehat{S}_{n}^{Airship} - \widetilde{S}_{n}^{BS}$;\\		
			\If{ $\widehat{S}_{n}^{Airship} \neq \emptyset $}			
			{				
				\For{ flow $i(1 \leq i \leq |\widehat{S}_{n}^{Airship}|)$ }
				{
					$r_{n}^{S} = r_{n}^{S} - E_{n}^{Satellite}$; \\
					\If{$r_{n}^{S} \geq 0$ }
					{
						$r_{n}^{A} = r_{n}^{A} - E_{n}^{Satellite}-E_{n}^{Airship}$; \\
						\If{$r_{n}^{A} \geq 0$ }
						{
							$u_{i,n}Q_{s_{i,n},r_{i,n}}=u_{i,n}$;\\
							$\widetilde{S}_{n}^{Airship} = \widetilde{S}_{n}^{Airship} \cup \widehat{S}_{n}^{Airship}(i)$;
						}	
						\Else
						{
							Inverse process for line $17$ and $19$;
						}				
						
					}
					\Else
					{
						Inverse process for line $16$;
					}
				}
			}	
			$\widehat{S}_{n}^{Satellite} =  \widehat{S}_{n}^{Satellite}  - \widetilde{S}_{n}^{BS}-\widetilde{S}_{n}^{Airship}$;\\
			\If{ $\widehat{S}_{n}^{Satellite} \neq \emptyset $}
			{				
				\For{ flow $i(1 \leq i \leq |\widehat{S}_{n}^{Satellite}|)$ }
				{
					$r_{n}^{S} = r_{n}^{S} - E_{n}^{Satellite}$; \\
					\If{$r_{n}^{S} \geq 0$ }
					{
						$u_{i,n}Q_{s_{i,n},r_{i,n}}=u_{i,n}$;\\
						$\widetilde{S}_{n}^{Satellite} = \widetilde{S}_{n}^{Satellite} \cup \widehat{S}_{n}^{Satellite}(i)$;
					}
					\Else
					{
						Inverse process for line $30$;
					}
				}	
			}			
			$o=o+1$;		
		}
		\Else
		{
			Enter Algorithm \ref{2-2};					
		}		
	}
\end{algorithm}

Additionally, at high train speeds, it becomes challenging to achieve link switching for every frame in practical applications \cite{linkswitch}. Therefore, we consider extending the link switching period. As the link switching period increases, the previous link selection and TS scheduling scheme may not be optimal for the flows in the current frame. For example, when the link switching period is $T_n=2T^{frame}$, Algorithm \ref{2-1} will allocate the optimal links and TSs for the flows in the first frame. However, if the allocation scheme from the first frame is still used for the flows in the second frame, the number of successfully scheduled flows may decrease due to time-varying channels and the lack of link switching. 

Then we propose Algorithm \ref{2-2} to address the above issue and maximize the weighted sum completed flows with frame $n\neq1+oK$ and $K>1$. To achieve this, we reorder the set of the flows for the BS-MR, satellite-airship-MR and satellite-MR links. The ordering is done in such a way that flows with high priority are preferred at the front, and flows with the same priority are sorted according to (\ref{w}). As a result, the flows in the sets $\widetilde{S}_{n}^{BS_j} $, $\widetilde{S}_{n}^{Airship}$ and $\widetilde{S}_{n}^{Satellite}$ are reordered into sets $\widehat{S}_{n}^{BS_j,sort} $, $\widehat{S}_{n}^{Airship,sort}$ and $\widehat{S}_{n}^{Satellite,sort}$. The number of occupied TSs for the flows in the sets $\widehat{S}_{n}^{BS_j,sort} $, $\widehat{S}_{n}^{Airship,sort}$ and $\widehat{S}_{n}^{Satellite,sort}$ are recorded in the sets $E_{n}^{BS_j}$, $E_{n}^{Airship}$ and $E_{n}^{Satellite}$
, respectively. If the set $\widetilde{S}_{n}^{BS_j} \neq \emptyset$, it indicates that some flows require the $j$-th BS scheduling. When the remaining TS of the $j$-th BS is $r_{n}^{BS_j} - E_{n}^{BS_j} \geq 0$, it means that the $j$-th BS can complete the QoS requirement of flow $i$ in the $n$-th frame; otherwise, the $j$-th BS cannot complete the QoS requirement of flow $i$ in the $n$-th frame. If the set $\widetilde{S}_{n}^{Airship} \neq \emptyset$, it indicates that some flows need to be co-scheduled by satellite and airship. When the remaining TS of the satellite and the airship is $r_{n}^{S} - E_{n}^{Satellite} \geq 0$ and $r_{n}^{A} - E_{n}^{Satellite}-E_{n}^{Airship} \geq 0$, it means that the satellite-airship-MR link can complete the QoS requirement of flow $i$ in the $n$-th frame. If the set $\widehat{S}_{n}^{Satellite} \neq \emptyset$, it indicates that some flows need to be scheduled by satellite. When the remaining TS of the satellite is $r_{n}^{S} - E_{n}^{Satellite} \geq 0$, it means that the satellite-MR link can complete the QoS requirement of flow $i$ in the $n$-th frame.

\begin{algorithm}[!t]
	\renewcommand{\thealgocf}{2-2}
	\caption{MWFS Algorithm with Frame $n\neq1+oK$}
	\label{2-2}
	\SetKwData{Or}{\textbf{or}}
	\DontPrintSemicolon
	%\SetAlgoLined
	\For{BS j($1 \leq j \leq 3$) }
	{
		\If{$\widetilde{S}_{n}^{BS_j} \neq \emptyset$}
		{
			\For{ flow $i(1 \leq i \leq |\widetilde{S}_{n}^{BS_j}|)$ }
			{
				$r_{n}^{BS_j} = r_{n}^{BS_j} - E_{n}^{BS_j}$; \\				
				\If{$r_{n}^{BS_j} \geq 0$ }
				{				
						$u_{i,n}Q_{s_{i,n},r_{i,n}}=u_{i,n}$;\\
					\Else
					{	
						Inverse process for line $4$;
					}				
	
				}
			}
		}
	}
	\If{$\widetilde{S}_{n}^{Airship} \neq \emptyset$}
	{
		\For{ flow $i(1 \leq i \leq |\widetilde{S}_{n}^{Airship}|)$ }
		{
				$r_{n}^{S} = r_{n}^{S} - E_{n}^{Satellite}$; \\
				\If{$r_{n}^{S} \geq 0$ }
				{
					$r_{n}^{A} = r_{n}^{A} - E_{n}^{Satellite}-E_{n}^{Airship}$; \\
					\If{$r_{n}^{A} \geq 0$ }
					{
						$u_{i,n}Q_{s_{i,n},r_{i,n}}=u_{i,n}$;\\
					}	
					\Else
					{
						Inverse process for line $11$ and $13$;
					}			
				}
				\Else
				{
					Inverse process for line $11$;
				}				
		}
	}
	\If{$\widetilde{S}_{n}^{Satellite} \neq \emptyset$}
	{
		\For{ flow $i(1 \leq i \leq |\widetilde{S}_{n}^{Satellite}|)$ }
		{
			$r_{n}^{S} = r_{n}^{S} - E_{n}^{Satellite}$; \\
			\If{$r_{n}^{S} \geq 0$ }
			{
						$u_{i,n}Q_{s_{i,n},r_{i,n}}=u_{i,n}$;\\
			}	
			\Else
			{	
				Inverse process for line $22$;
			}				
		}
	}	
\end{algorithm}

For the complexity of the Algorithm \ref{2-1}, we can observe the
outer for loop has $N^{frame}$ iterations. The inner for loop has $(3|\widehat{S}_{n}^{BS_j}| + |\widehat{S}_{n}^{Airship}| + |\widehat{S}_{n}^{Satellite}|)$ iterations. For the complexity of Algorithm \ref{2-2}, it has $(3|\widetilde{S}_{n}^{BS_j}| + |\widetilde{S}_{n}^{Airship}| + |\widehat{S}_{n}^{Satellite}|)$ iterations. For the worst-case, the complexity for MWFS algorithm is $\mathcal O (N^{frame}(3|\widehat{S}_{n}^{BS_j}| + |\widehat{S}_{n}^{Airship}| + |\widehat{S}_{n}^{Satellite}|))$.

\section{PERFORMANCE EVALUATION}
In this section, we evaluate the performance of the proposed algorithm based on the weighted sum completed flows and the total transmitted bits with other benchmark algorithms for the HSR mmWave communication in SAGIN. 

\subsection{Simulation Setup}
Consider a train moving sequentially through the coverage areas of three BSs. To address potential coverage blind spots between these BSs, an airship and a satellite are deployed to supplement the coverage, ensuring continuous connectivity throughout the train's journey. The Rician factor $K=10$ dB for BS-MR link. The Rician factors for satellite-airship-MR and satellite-MR links rely on the specific elevation angle \cite{3gpp.38.811}. The coordinates of BS$_1$, BS$_2$ and BS$_3$ are $O_1(-7\times10^3,10,15)$, $O_2(-1.2\times10^3,10,15)$ and  $O_2(1.2\times10^3,10,15)$. It is important to note that the above coordinates are in meters. The initial latitude, longitude and altitude of the satellite are obtained based on Satellite Tool Kit (SKT). Since the latitude and longitude of the MRs hardly change as the train travels in the limited time, we assume that the latitude and longitude of the MRs are the same as those of the airship. The inclination of the satellite is $90^\circ$, meaning that its orbit is perpendicular to the Earth's equatorial plane and passes over both poles. The gains of transmitted antenna for BSs, airship and satellite are $G^{BS}_t=6$ dBi, $G^{A}_t=20$ dBi and $G^{S}_t=38.5$ dBi, respectively \cite{3gpp.38.821}. The bandwidth 1 is set to $W_1=400$ MHz for the satellite-airship link and satellite-MR link. The bandwidth 2 is set to $W_2=850$ MHz for the BS-MR link and airship-MR link \cite{3gpp.38.104}. The $x$-coordinate of the MRs will be updated at each frame. We set the MRs move $3$ m at each frame by adjusting the the number of TSs in the frame. The QoS requirements of flows are from $10$ to $3000$ Mbps, which correspond to $QoS \times T^{frame}$ for a frame. The specific parameters are shown in Table \ref{pa} \cite{wangyibing},\cite{Niu}.

\begin{table}[t]
	\captionsetup{labelsep=none}
	\captionsetup{font=footnotesize}
	\caption{\\SIMULATION PARAMETERS}
	\centering
	\begin{tabular}{lll}
	\hline
	Parameter & Symbol & Value \\
	\hline
	Frequency band & $f_c$  & $28$ GHz  \\
	Bandwidth 1 & $W_1$  & $400$ MHz \\ 
	Bandwidth 2 & $W_2$  & $850$ MHz \\ 
	Number of the MRs & $N^{MR}$ & $16$ \\
	Coordinates of BS$_1$     & $O_1$  & $(-7\times10^3,10,15)$ \\
	Coordinates of BS$_2$     & $O_2$  & $(-1.2\times10^3,10,15)$ \\
	Coordinates of BS$_3$     & $O_3$  & $(1.2\times10^3,10,15)$ \\
	Height of MRs     & $h^{MR}$  & $4$ m \\			
	Distance of between MRs & $d^{MR}$  & $30$ m \\
	Rician factor for BS-MR link & $K$  &  $10$ dB \\
	Transmitted power of satellite & $P_t^{S}$  & $60$ dBm \\
	Transmitted power of airship & $P_t^{A}$  & $30$ dBm \\
	Transmitted power of BSs & $P_t^{BS}$  & $30$ dBm \\
	Transmitted antenna gain of satellite & $G^{S}_t$  & $38.5$ dBi \\
	Transmitted antenna gain of airship & $G^{A}_t$  & $20$ dBi \\
	Transmitted antenna gain of BSs & $G^{BS}_t$  & $6$ dBi \\
	Received antenna gain of airship & $G^{A}_r$  & $20$ dBi \\
	Received antenna gain of MRs & $G^{MR}_r$  & $6$ dBi \\
	Atmospheric attenuation & $l_A$  & $0.1$ dB/km \\
	Speed of satellite & $v_s$  & $7.5$ km/s \\
	Received sensitivity & $R_s$  & $-80$ dBm \\	
	Transceiver efficiency factor & $\varepsilon$ & $0.5$ \\
	Slot time & $\Delta t$ & $18$ $\mu$s \\
	Beacon period & $T_s$ & $850$ $\mu$s \\
	QoS requirements of flows & $QoS$ & $10\sim3000$ Mbps \\	
	\hline	
	\end{tabular}
	\label{pa}
\end{table}
To demonstrate the performance enhancements of the proposed algorithm, we conducted comparisons with five algorithms. These include the BS-MR link only (BLO) algorithm, the satellite-airship-MR link only (SLO) algorithm, the satellite-airship-MR and satellite-MR links (SASL) algorithm, the MFS algorithm, and the random communication (RC) algorithm. The specific details are as follows:
\begin{itemize}
	\item[1)] \emph{BLO}: Only use terrestrial BSs to schedule flows for MRs according to the QoS requirements of flows.	
	\item[2)] \emph{SLO}: The satellite-airship-MR link schedule flows for MRs according to the QoS requirements of flows.
	\item[3)] \emph{SASL}: Both satellite-airship-MR and satellite-MR links schedule flows for MRs according to the QoS requirements of flows.
	\item[4)] \emph{MFS} \cite{ll}: The MFS algorithm consists of relay selection and flow transmission scheduling, where the flow transmission scheduling is designed in a QoS-aware manner. However, it does not take the priority of flows and the satellite-MR link into account.
	\item[5)] \emph{RC}: All links can schedule flows for MRs according to the QoS requirements, but these links do not optimize the order of the scheduling flows.
\end{itemize}

For above benchmark algorithms, flows with high-priority are transmitted first. Besides, two performance metrics considered are:
\begin{itemize}
	\item[1)] \emph{Weighted sum completed flows}: The sum of the weight coefficients of the flows that are scheduled successfully.
	\item[2)] \emph{Total transmitted bits}: The total number of bits sent for MRs by the transmitters, where the transmitters include the airship, satellite and BSs.	
\end{itemize}

\subsection{Comparison With Other Algorithms}
1) In the case of different time: We set that the time is $0 \sim 3,600$ frames, where the first frame indicates the first MR enters the BS$_1$'s coverage area. The height of airship is set to $h^A = 2$ km. The satellite height is set to $h^S=600$ km. The train speed is set to $v_t=500$ km/h. The period of the link switching is set to $T_p = T^{frame}$. The elevation angle between the MR and the satellite is $90^\circ$ in the first frame.

\begin{figure}[t]
	\centering
	\begin{minipage}[c]{0.5\textwidth}
		\centering
		\includegraphics[width=3.3in]{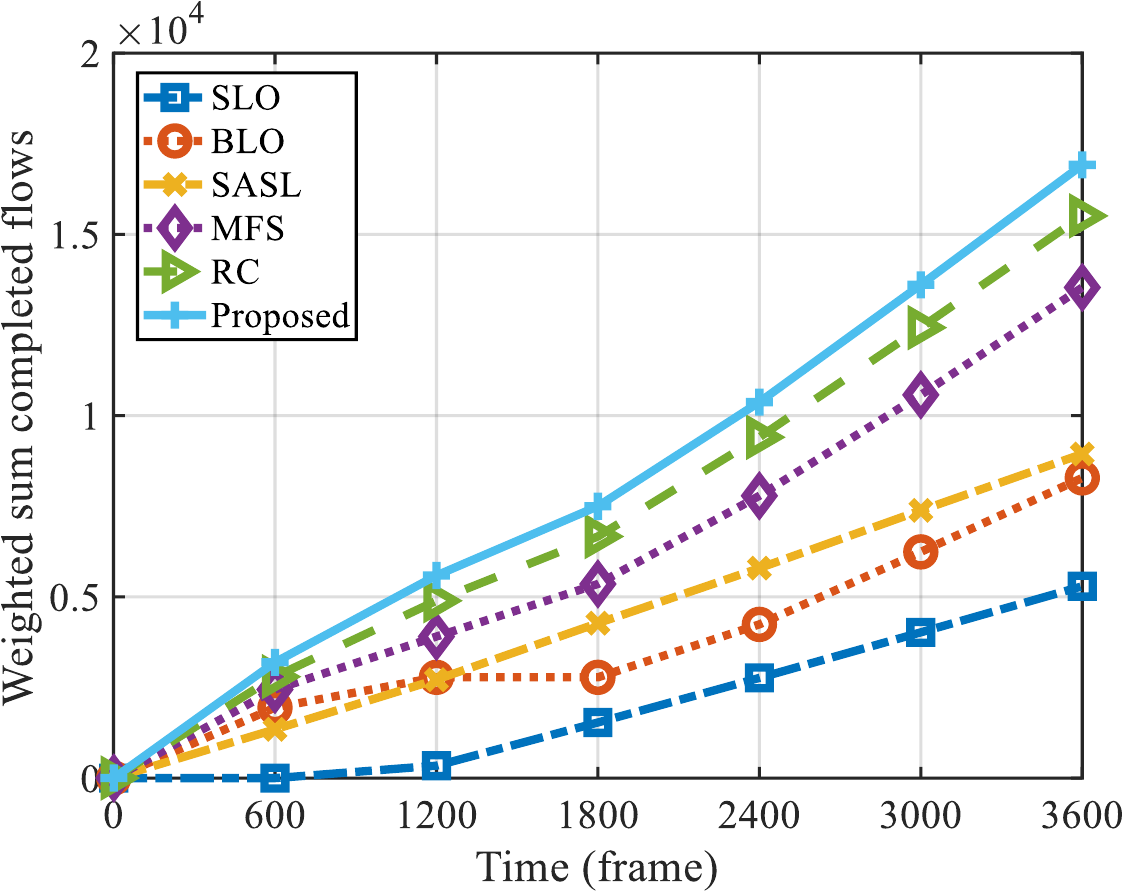}\\
		\caption{Weighted sum completed flows versus different time.}
		\label{time_flow}
	\end{minipage} \\
\end{figure}

Fig. \ref{time_flow} illustrates the weighted sum completed flows over time. As time progresses, the weighted sum completed flows increase for all six algorithms. The proposed algorithm outperforms the others by completing the weighted flows. Closely following, the RC algorithm achieves the second-highest completion rate, attributed to its ability to serve multiple requests simultaneously through the BS-MR, satellite-airship-MR, and satellite-MR links. However, the RC algorithm's random assignment for flows may lead to link blockage, particularly when handling flows with large QoS requirements, which could prevent other flows from being scheduled. The MFS algorithm ranks third, which does not consider the satellite-MR link and the priority of flows. Lastly, the comparison between the SASL and SLO algorithms reveals that the satellite-MR link can enhance the completion of weighted flows. 

\begin{figure}[t]
	\centering
	\begin{minipage}[c]{0.5\textwidth}
		\centering
		\includegraphics[width=3.3in]{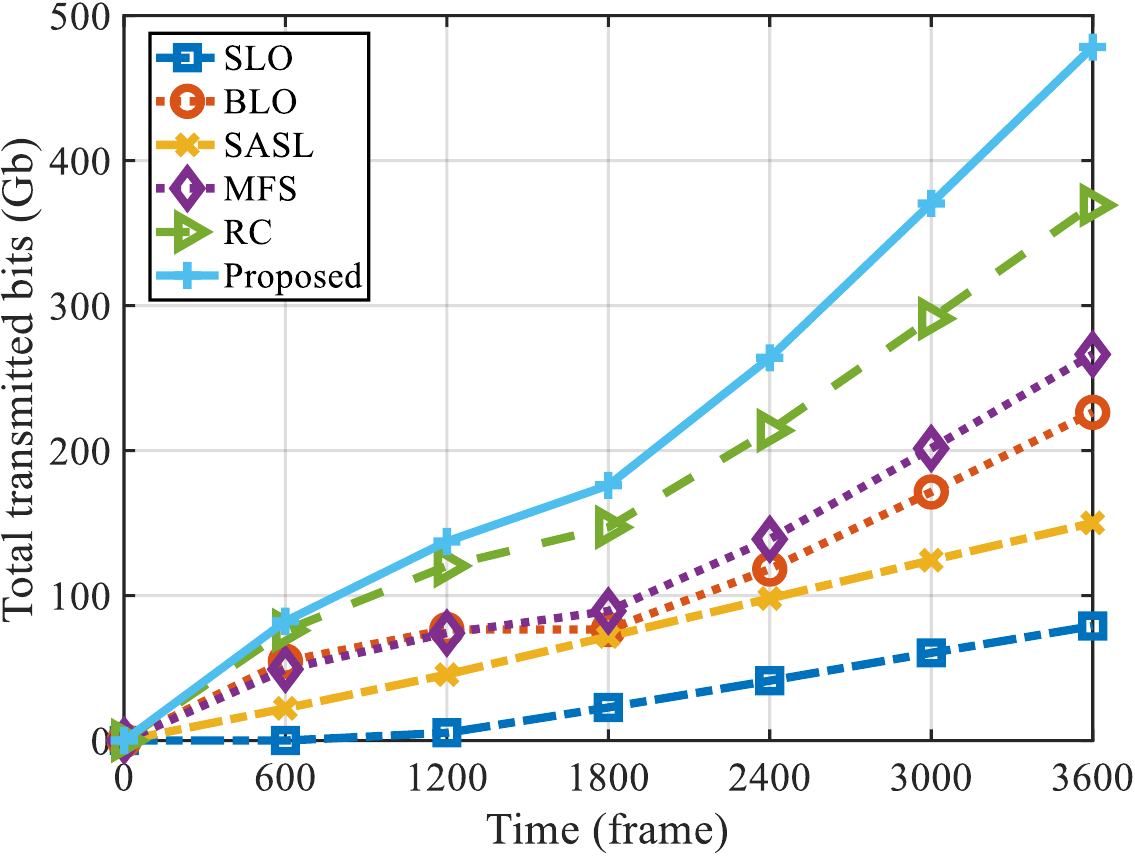}\\
		\caption{Total transmitted bits versus the different time.}
		\label{flow_bit}
	\end{minipage} \\
\end{figure}

Fig. \ref{flow_bit} illustrates the total transmitted bits over time. The results indicate that the total transmitted bits increase for all six algorithms as time progresses. Moreover, the proposed algorithm performs significantly better than the others. Additionally, the RC algorithm transmits more bits than the MFS, BLO, SASL and SLO algorithms. This suggests that the simultaneous use of the BS-MR, satellite-airship-MR, and satellite-MR links can significantly boost bit transmission for HSR mmWave communication in SAGIN, even without optimizing the flow scheduling order.

2) In the case of different speed of the train: We set that the train speed $v_t$ is from $100$ to $500$ km/h. The duration is $3,600$ frames. The height of airship is set to $h^A = 2$ km. The satellite height is set to $h^S=600$ km. The period of the link switching is set to $T_p = T^{frame}$. The elevation angle between the MR and the satellite is $90^\circ$ in the first frame.

\begin{figure}[t]
	\centering
	\begin{minipage}[c]{0.5\textwidth}
		\centering
		\includegraphics[width=3.3in]{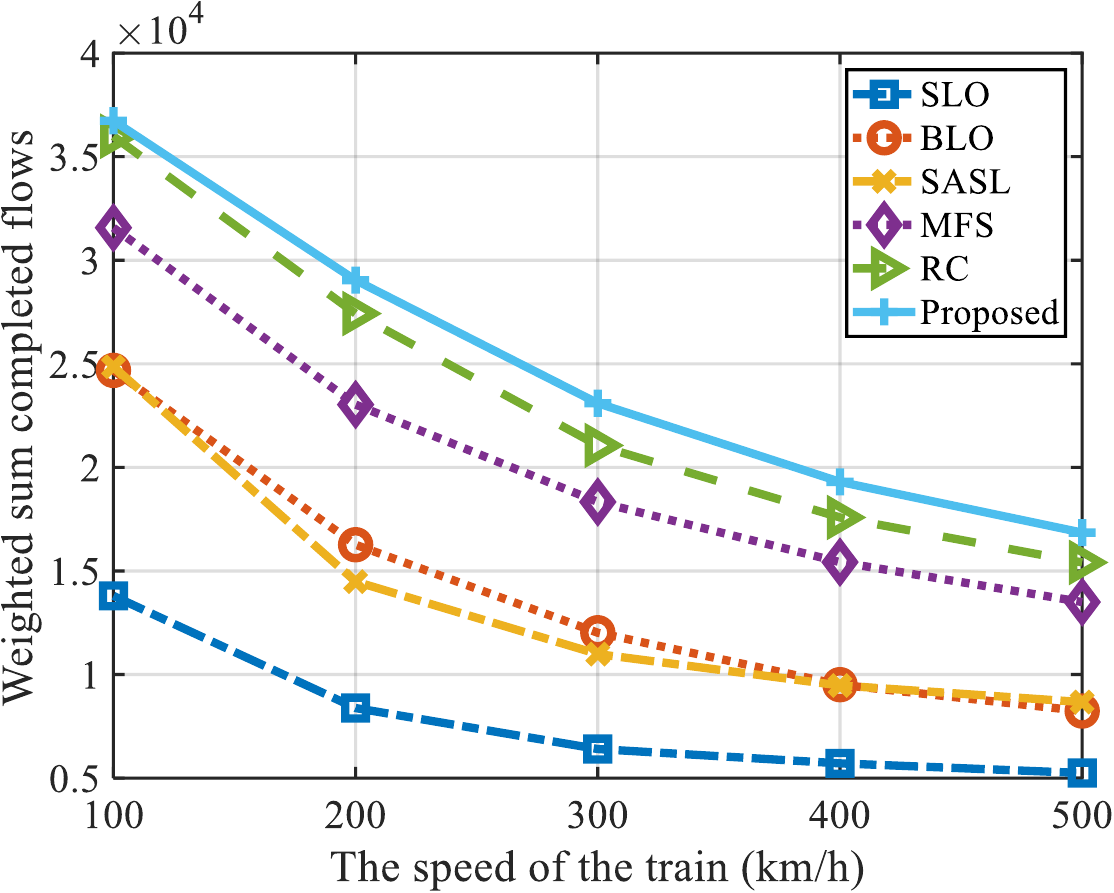}\\
		\caption{Weighted sum completed flows versus different speed of the train.}
		\label{v_flow}
	\end{minipage} \\
\end{figure}

Fig. \ref{v_flow} illustrates the weighted sum completed flows changes with the train speed. Firstly, it can be seen that as the train speed increases, the weighted sum completed flows for all six algorithms gradually decreases. This is because higher train speeds reduce the communication time for MRs over a finite distance. Secondly, the proposed algorithm completes the maximum number of weighted flows at different speeds, which indicates that the proposed method is suitable for the HSR scenario. Finally, at a speed of $100$ km/h, the proposed algorithm does not significantly outperform the RC algorithm. This is because, at lower speeds, the number of TSs in each frame is large enough to easily satisfy the QoS requirements for both the proposed and RC algorithms.

\begin{figure}[t]
	\centering
	\begin{minipage}[c]{0.5\textwidth}
		\centering
		\includegraphics[width=3.3in]{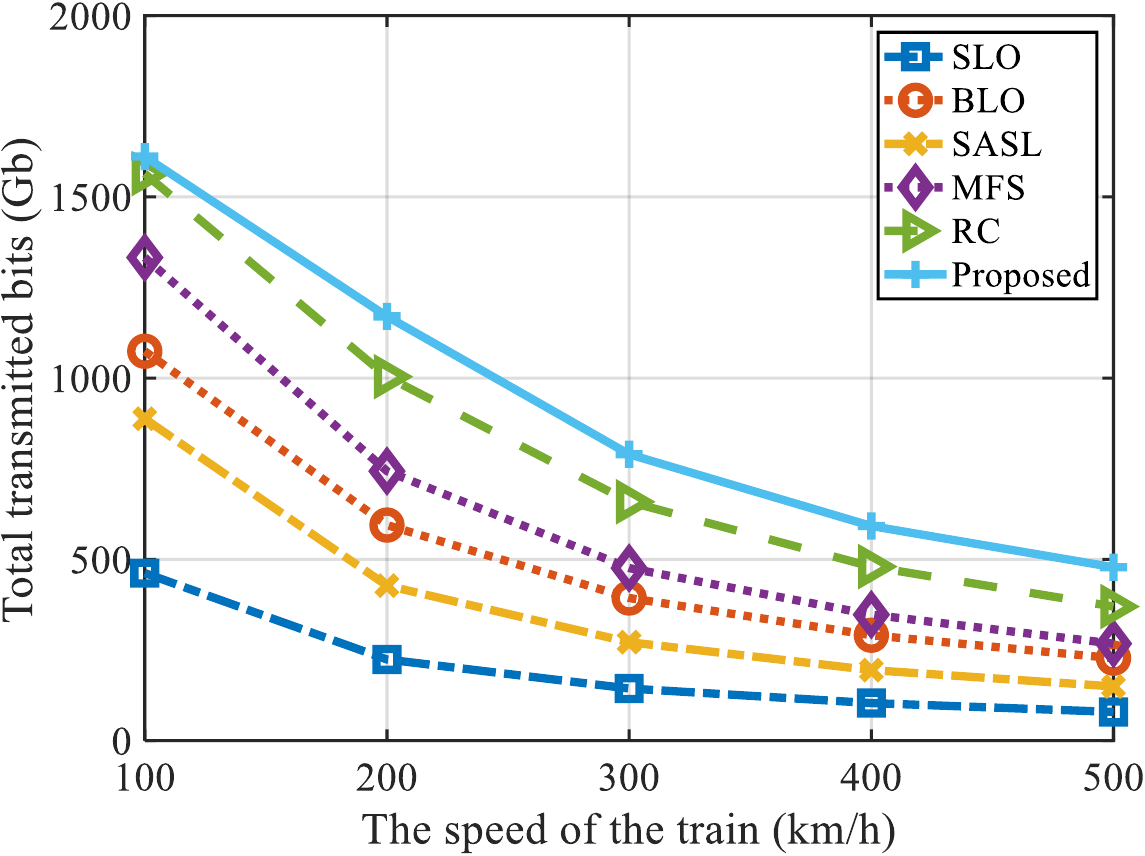}\\
		\caption{Total transmitted bits versus different speed of the train.}
		\label{v_bit}
	\end{minipage} \\
\end{figure}

Fig. \ref{v_bit} illustrates the total transmitted bits changes with the train speed. The proposed algorithm transmits the most bits, with a performance 29\% higher than the RC algorithm at a train speed of 500 km/h. Additionally, the SLO algorithm performs very poorly in terms of transmitted bits at high speeds.

3) In the case of different periods of the link switching: The period of the link switching is set to $T^{frame}\leq T_p\leq10T^{frame}$. The duration is $3,600$ frames. The height of airship is set to $h^A=2$ km. The satellite height is set to $h^S=600$ km. The train speed is set to $v_t=500$ km/h. The elevation angle between the MR and the satellite is $90^\circ$ in the first frame.

\begin{figure}[t]
	\centering
	\begin{minipage}[c]{0.5\textwidth}
		\centering
		\includegraphics[width=3.3in]{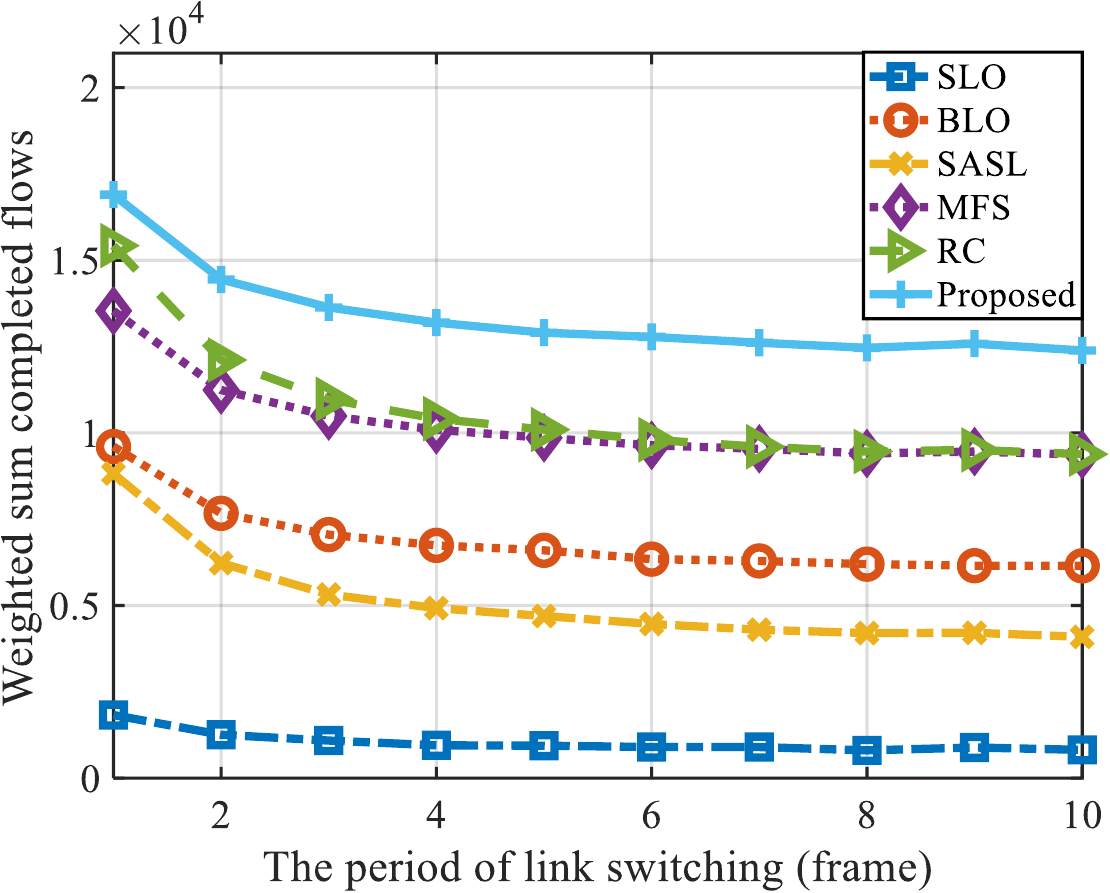}\\
		\caption{Weighted sum completed flows versus different periods of link switching.}
		\label{period_flow}
	\end{minipage} \\
\end{figure}

Fig. \ref{period_flow} illustrates the weighted sum completed flows as the link switching period varies. Notably, the weighted sum completed flows for all six algorithms decrease as the period increases. The main reason is that it becomes difficult to select an appropriate MR for the satellite, airship, and BS to allocate TSs due to a lack of timeliness. Moreover, the proposed algorithm outperforms the RC, MFS, BLO, SASL, and SLO algorithms in weighted sum completed flows, even at $10T^{frame}$.

\begin{figure}[t]
	\centering
	\begin{minipage}[c]{0.5\textwidth}
		\centering
		\includegraphics[width=3.3in]{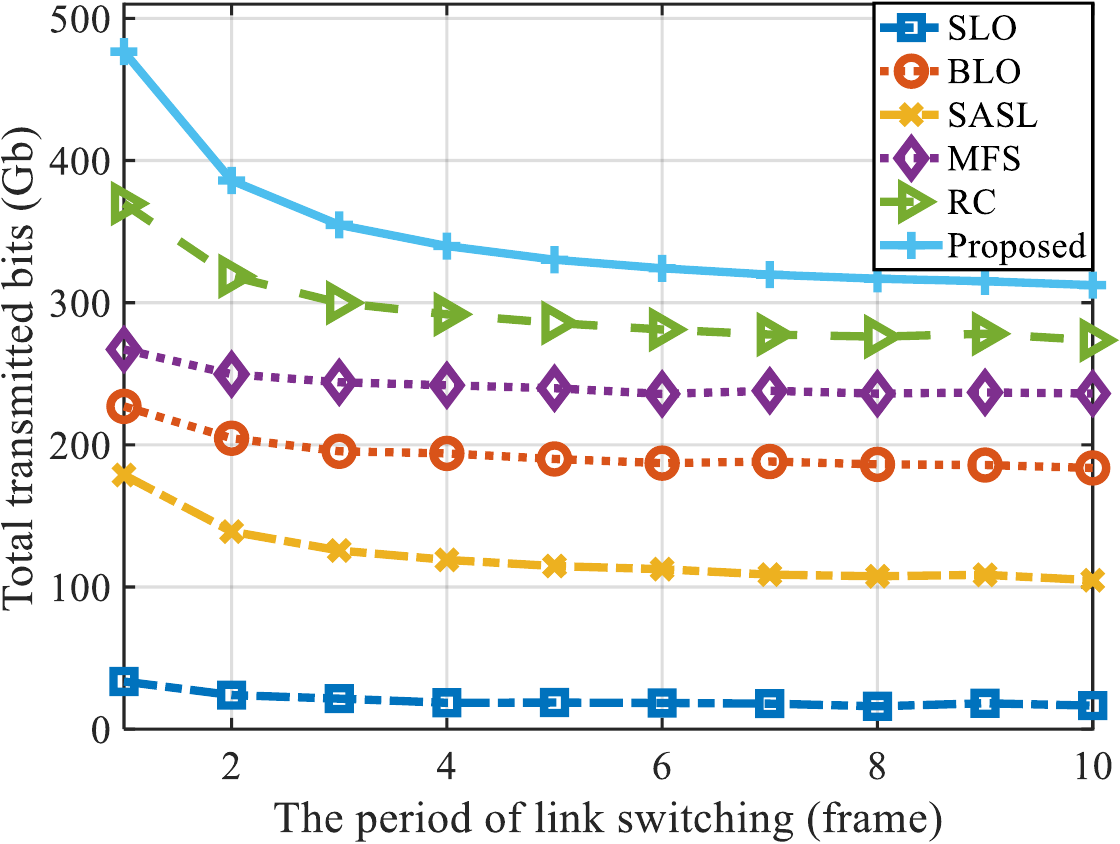}\\
		\caption{Total transmitted bits versus different periods of link switching.}
		\label{period_bit}
	\end{minipage} \\
\end{figure}

Fig. \ref{period_bit} illustrates the total transmitted bits as the link switching period varies. As the link switching period increases, there is a noticeable decrease in the total transmitted bits. However, the proposed algorithm consistently outperforms the others, maintaining the highest number of transmitted bits regardless of the link switching period. This demonstrates the effectiveness of the proposed algorithm even at longer link switching period.

4) In the case of different heights of airship: We set the height of airship is from $1$ to $4$ km. The duration is $3,600$ frames. The satellite height is set to $h^S=600$ km. The train speed is set to $v_t=500$ km/h. The period of the link switching is set to $T_p = T^{frame}$. The elevation angle between the MR and the satellite is $90^\circ$ in the first frame.

\begin{figure}[t]
	\centering
	\begin{minipage}[c]{0.5\textwidth}
		\centering
		\includegraphics[width=3.3in]{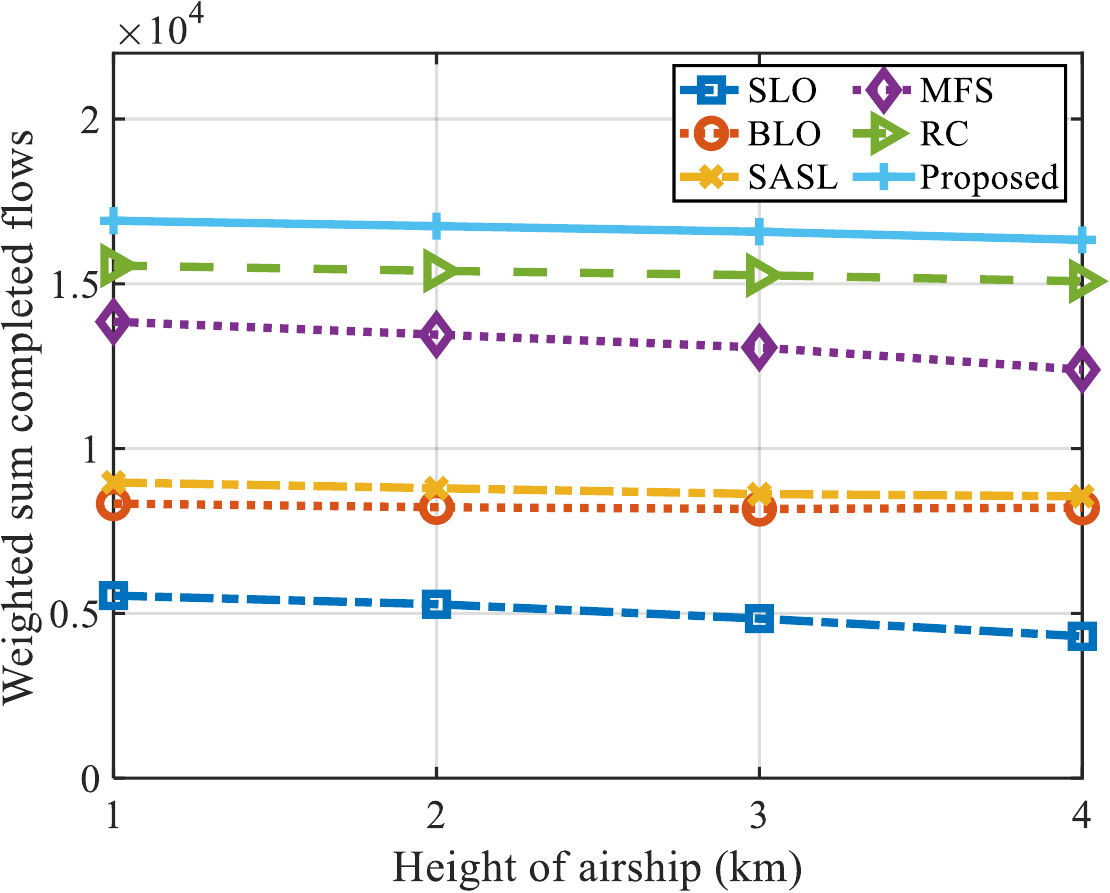}\\
		\caption{Weighted sum completed flows versus different heights of airship.}
		\label{h_flow}
	\end{minipage} \\
\end{figure}

Fig. \ref{h_flow} illustrates the weighted sum completed flows changes with the airship's height. We can see that the weighted sum completed flows gradually decreases as the airship's height increases for the proposed, RC, MFS, SASL and SLO algorithms. The main reason is that an increase in the airship’s height causes the distance between the airship and MR to increase, leading to a decrease in the transmission rate of the airship-MR link. Although a higher airship’s height reduces the distance between the airship and the satellite, it is evident that the transmission rate of the airship-MR link has a greater impact on system performance than the satellite-airship link. Additionally, the BLO algorithm is unaffected by the airship’s height.

\begin{figure}[t]
	\centering
	\begin{minipage}[c]{0.5\textwidth}
		\centering
		\includegraphics[width=3.3in]{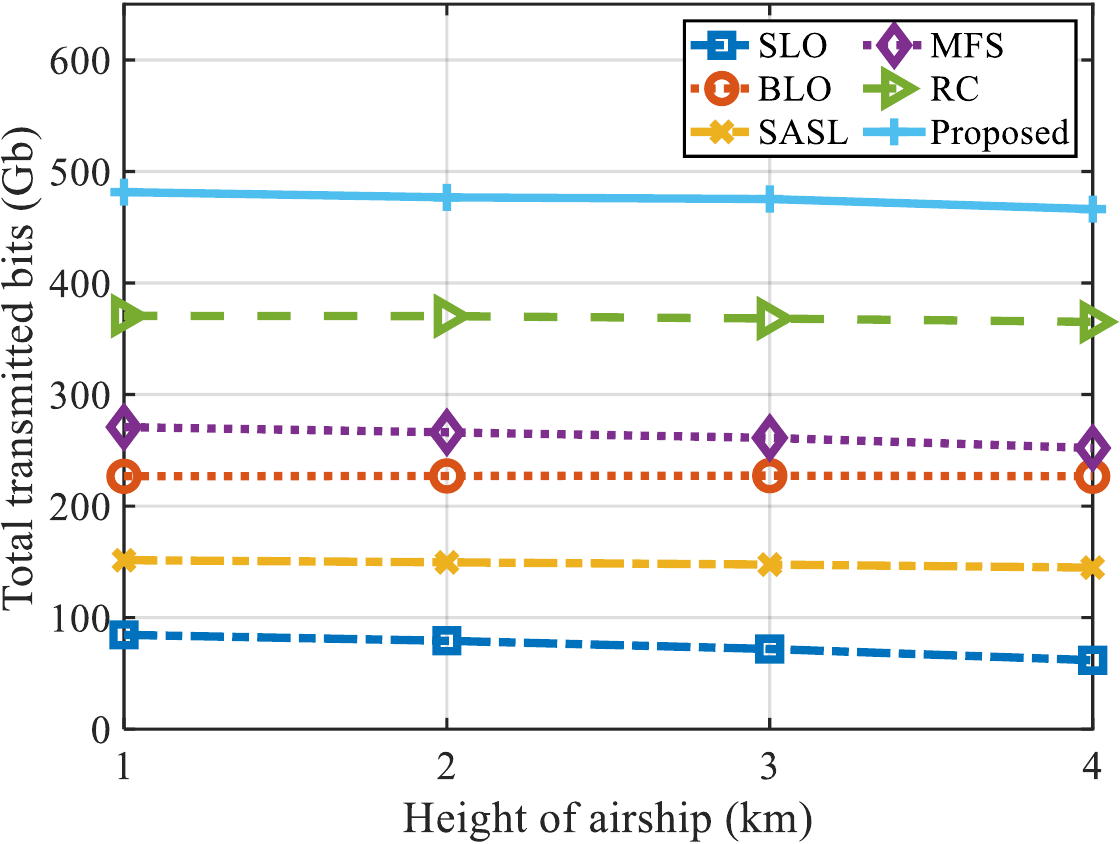}\\
		\caption{Total transmitted bits versus different heights of airship.}
		\label{h_bit}
	\end{minipage} \\
\end{figure}

Fig. \ref{h_bit} illustrates the total transmitted bits versus different heights of airship. For the proposed, RC, MFS, SASL, and SLO algorithms, the total transmitted bits slightly decreases with increasing airship's height. The reason is that when the airship's height increases, the distance between the airship and the MRs becomes greater, resulting in a decrease in the transmission rate. Notably, the proposed algorithm consistently transmits the most bits at different heights of the airship, outperforming the other algorithms.

5) In the case of different heights of satellite: We set that the height of satellite is from $300$ to $1500$ km. The duration is $3,600$ frames. The height of airship is set to $h^A=2$ km. The train speed is set to $v_t=500$ km/h. The period of the link switching is set to $T_p = T^{frame}$. The elevation angle between the MR and the satellite is $90^\circ$ in the first frame.

\begin{figure}[t]
	\centering
	\begin{minipage}[c]{0.5\textwidth}
		\centering
		\includegraphics[width=3.3in]{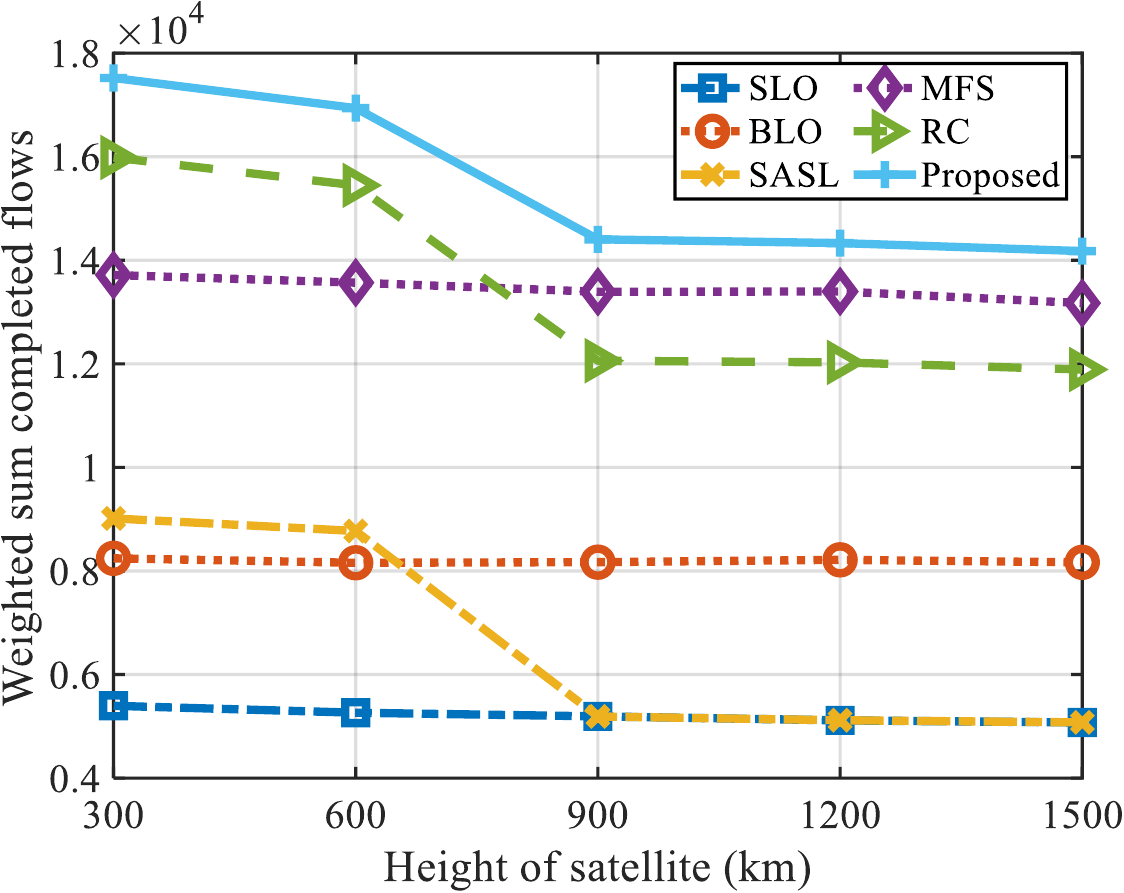}\\
		\caption{Weighted sum completed flows versus different heights of satellite.}
		\label{hs_flow}
	\end{minipage} \\
\end{figure}

Fig. \ref{hs_flow} illustrates the weighted sum completed flows versus different heights of satellite. As the satellite’s height increases, we can see that the weighted sum completed flows decrease for the proposed, RC and SASL algorithms. Especially when the satellite's height exceeds 600 km, the performance of the proposed, RC, and SASL algorithms declines significantly. The main reason is that as the distance between the satellite and the ground increases, it becomes more difficult to meet the QoS requirements of flows in the satellite-MR link. Additionally, the MFS and the SLO algorithms are scarcely affected by the satellite’s height, indicating that the satellite-airship-MR link can still satisfy the QoS requirements of flows. 

\begin{figure}[t]
	\centering
	\begin{minipage}[c]{0.5\textwidth}
		\centering
		\includegraphics[width=3.3in]{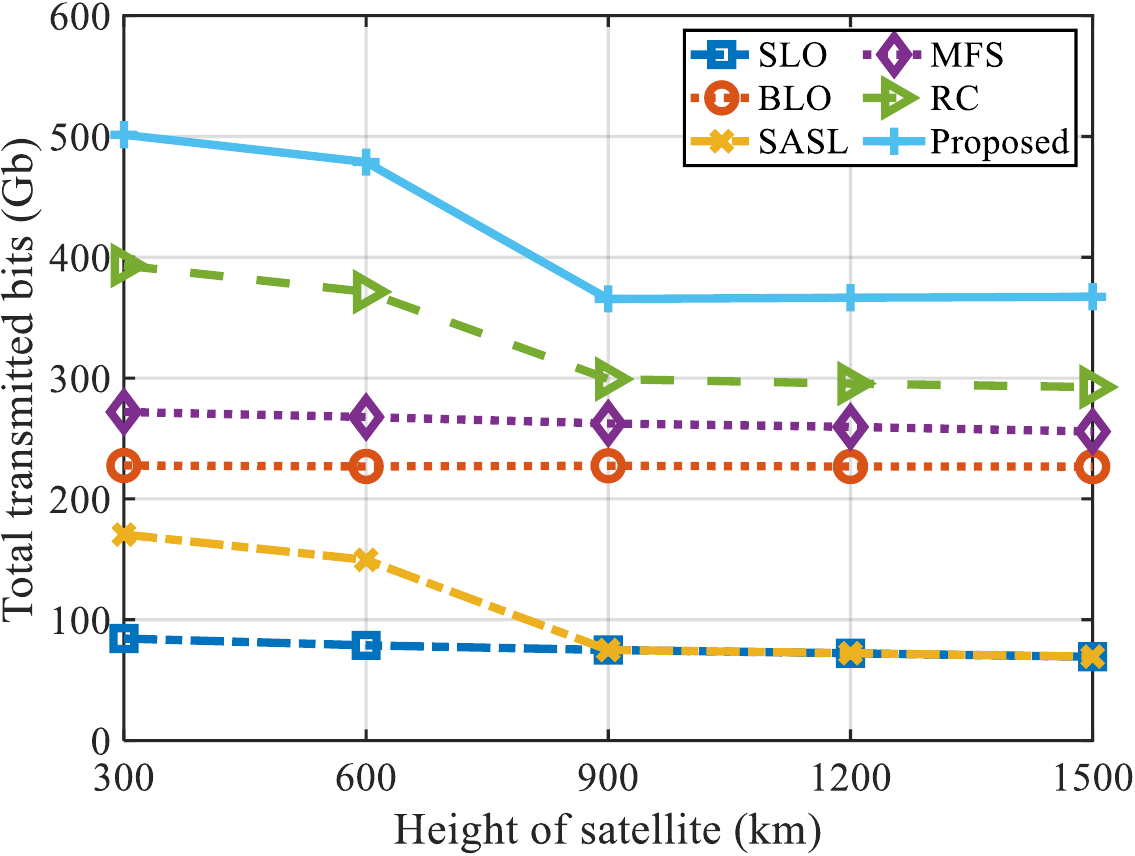}\\
		\caption{Total transmitted bits versus different heights of satellite.}
		\label{hs_bit}
	\end{minipage} \\
\end{figure}

Fig. \ref{hs_bit} illustrates the total transmitted bits versus different heights of satellite. The proposed algorithm consistently transmits the most bits. Notably, the satellite-MR link struggles to meet the QoS requirements of flows when the height of satellite is more than 900 km, resulting in a decrease in the total transmitted bits for the proposed, RC, and SASL algorithms.

6) In the case of different elevation angles between the MR and satellite: We set the elevation angle range from $10^\circ$ to $90^\circ$. The duration is $3,600$ frames. The height of airship is set to $h^A=2$ km. The satellite height is set to $h^S=600$ km. The train speed is set to $v_t=500$ km/h. The period of the link switching is set to $T_p = T^{frame}$.

\begin{figure}[t]
	\centering
	\begin{minipage}[c]{0.5\textwidth}
		\centering
		\includegraphics[width=3.3in]{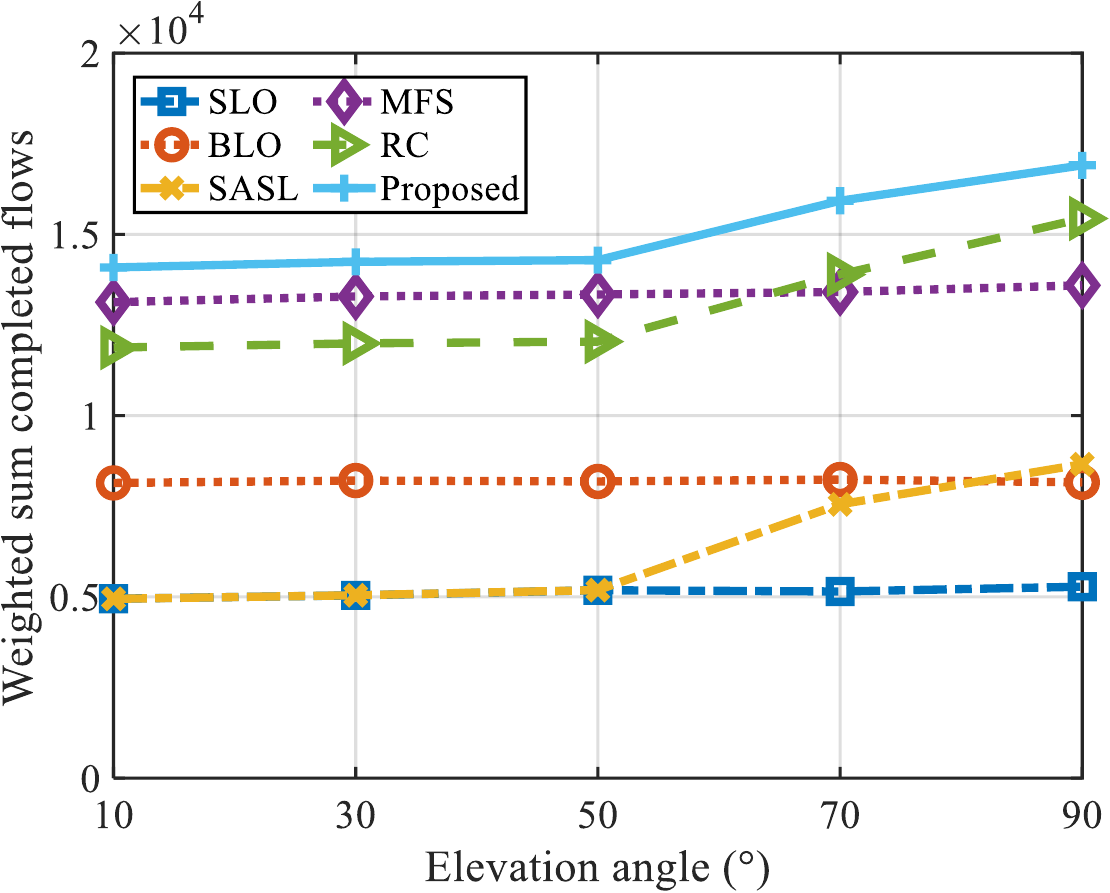}\\
		\caption{Weighted sum completed flows versus different elevation angles.}
		\label{sa_flow}
	\end{minipage} \\
\end{figure}

Fig. \ref{sa_flow} illustrates the weighted sum completed flows versus different elevation angles. As the elevation angle increases, the weighted sum completed flows significantly rise for the proposed, RC, and SASL algorithms. This is because an increase in the elevation angle reduces the distance between the satellite and the MR, allowing the satellite-MR link to more easily meet the QoS requirements for flows. Conversely, when the elevation angle is small, the satellite is too far from the MR, resulting in a reduction in the weighted sum completed flows. Notably, the proposed algorithm achieves the highest weighted sum completed flows, outperforming the others.

\begin{figure}[t]
	\centering
	\begin{minipage}[c]{0.5\textwidth}
		\centering
		\includegraphics[width=3.3in]{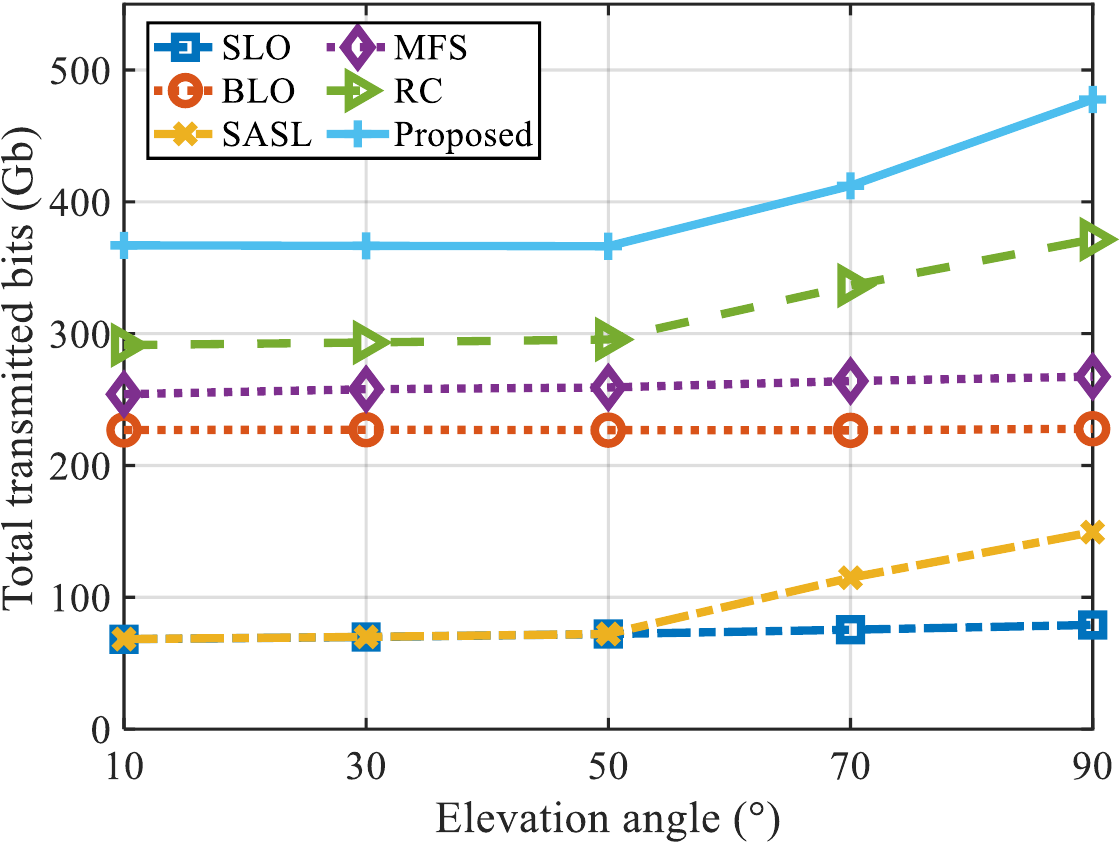}\\
		\caption{Total transmitted bits versus different elevation angles.}
		\label{sa_bit}
	\end{minipage} \\
\end{figure}

Fig. \ref{sa_bit} illustrates the total transmitted bits versus different elevation angles. Similarly, the proposed algorithm achieves the highest number of transmitted bits. When the elevation angle reaches $90^\circ$, the satellite is positioned directly above the airship and the MR, allowing both the satellite-airship-MR link and the satellite-MR link to function effectively. When the elevation less than $50^\circ$, the satellite is too far from the MR, which leads to the satellite-MR link struggling to meet the QoS requirements for flows.

\section{Conclusion}

In this paper, we address the transmission scheduling problem for HSR mmWave communications in SAGIN. First, we introduce an optimization problem aimed at maximizing the weighted sum completed flows that meet the QoS requirements. This optimization problem considers the movement of the train, the positions of the satellite, the QoS requirements of the flows, channel variations, the priority of the flows and the scheduling order of flows. Next, we propose a link selection algorithm for MRs to identify links that satisfy the QoS requirements of flows in each frame. Additionally, to increase the weighted sum completed flows, we propose a MWFS algorithm that considers the priority of different flows and scheduling order of flows. We also optimize the scheduling plan for flows in the different periods of link switching. Simulation results show that the proposed algorithm effectively improves the weighted sum completed flows and the total transmitted bits as the train passing through coverage areas of the BSs, airship, and satellite. Furthermore, we find that proposed algorithm can enhance system performance even with a higher height of the satellite or a long link switching period.

\footnotesize % 调了顺序后，参考文献字体变大，回归正常字体
\bibliographystyle{ieeetr} % 目前来看没什么用，可以删掉
\balance
\bibliography{liulei_arVix.bib} % 读取liulei_arVix.bib文件中的参考文献

\bibliographystyle{IEEEtran}
%\bibliography{IEEEabrv,Bibliography}
\vspace{-1cm}
\begin{IEEEbiography}[{\includegraphics[width=1in,height=1.25in,clip,keepaspectratio]{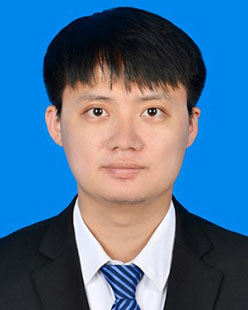}}]{Lei Liu}
	received the B.E. degree in electronic information engineering from Nantong University, Nantong, China, in 2017, and the M.S. degree in electronic and communication engineering from Inner Mongolia University, Hohhot, China, in 2021. He is currently  working toward the Ph.D. degree with the School of Electronic and Information Engineering, Beijing Jiaotong University, Beijing, China. His research includes millimeter-wave wireless communications, resource allocation and channel estimation.
\end{IEEEbiography}

\vspace{-1cm}

\begin{IEEEbiography}[{\includegraphics[width=1in,height=1.25in,clip,keepaspectratio]{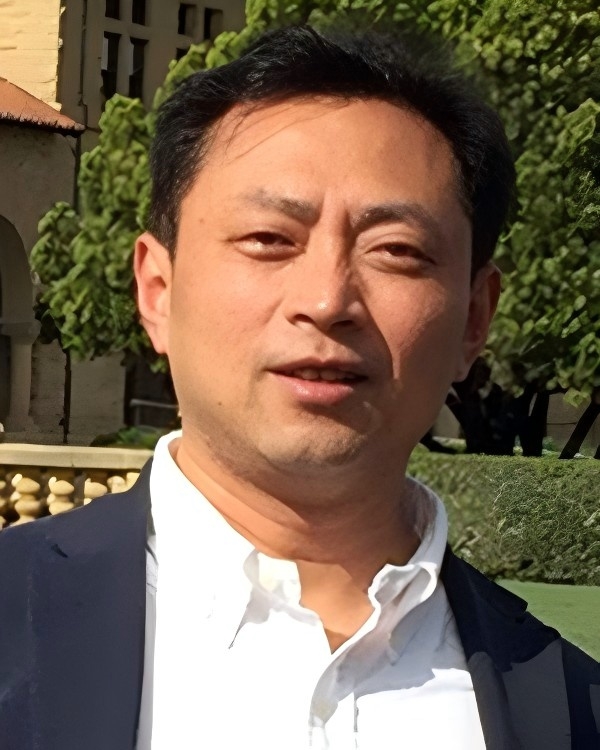}}]{Bo Ai}
	(Fellow, IEEE) received the M.S. and Ph.D. degrees from Xidian University, China. 
	
	He studies as a postdoctoral student with Tsinghua University. He was a Visiting Professor with the Electrical Engineering Department, Stanford University in 2015. He is currently with Beijing Jiaotong University as a Full Professor and a Ph.D. Candidate Advisor, where he is the Deputy Director	of the State Key Laboratory of Rail Traffic Control and Safety and the International Joint Research Center. He is one of the main people responsible	for the Beijing Urban Rail Operation Control System, International Science and Technology Cooperation Base. He is also a member of the Innovative Engineering Based jointly granted by the Chinese Ministry of Education and the State Administration of Foreign Experts Affairs. He was honored with the Excellent Postdoctoral Research Fellow by Tsinghua University in 2007.	He has authored/coauthored eight books and published over 300 academic	research papers in his research area. He holds 26 invention patents. He has been the research team leader for 26 national projects. His interests include the	research and applications of channel measurement and channel modeling and dedicated mobile communications for rail traffic systems. He has been notified by the Council of Canadian Academies that, based on Scopus database, he has been listed as one of the Top 1\% authors in his field all over the world. He has also been feature interviewed by the IET Electronics Letters. He has received some important scientific research prizes.

\end{IEEEbiography}

\vspace{-1cm}

\begin{IEEEbiography}[{\includegraphics[width=1in,height=1.25in,clip,keepaspectratio]{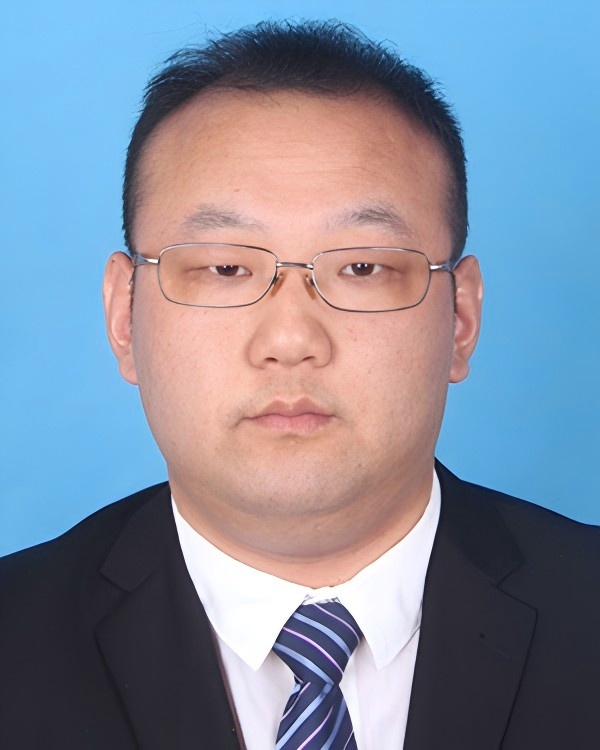}}]{Yong Niu}
	(Senior Member, IEEE) received the B.E. degree in electrical engineering from Beijing Jiaotong University, Beijing, China, in 2011, and the Ph.D. degree in electronic engineering from Tsinghua University, Beijing, China, in 2016. 
	
	From 2014 to 2015, he was a Visiting Scholar with the University of Florida, Gainesville, FL, USA. He is currently an Associate Professor with the State Key Laboratory of Advanced Rail Autonomous Operation, Beijing Jiaotong University. His research interests include networking and communications, including millimeter wave communications, device-to-device communication, medium access control, and software-defined networks. He was the recipient the Ph.D. National Scholarship of China in 2015, the Outstanding Ph.D. Graduates and Outstanding Doctoral Thesis of Tsinghua University in 2016, the Outstanding Ph.D. Graduates of Beijing in 2016, and the Outstanding Doctorate Dissertation Award from the Chinese Institute of Electronics in 2017. He was the Technical Program Committee Member for IWCMC 2017, VTC2018-Spring, IWCMC 2018, INFOCOM 2018, and ICC 2018. He was the Session Chair for IWCMC 2017. He was also the recipient of the 2018 International Union of Radio Science Young Scientist Award.
\end{IEEEbiography}
\vspace{-1cm}

\begin{IEEEbiography}[{\includegraphics[width=1in,height=1.25in,clip,keepaspectratio]{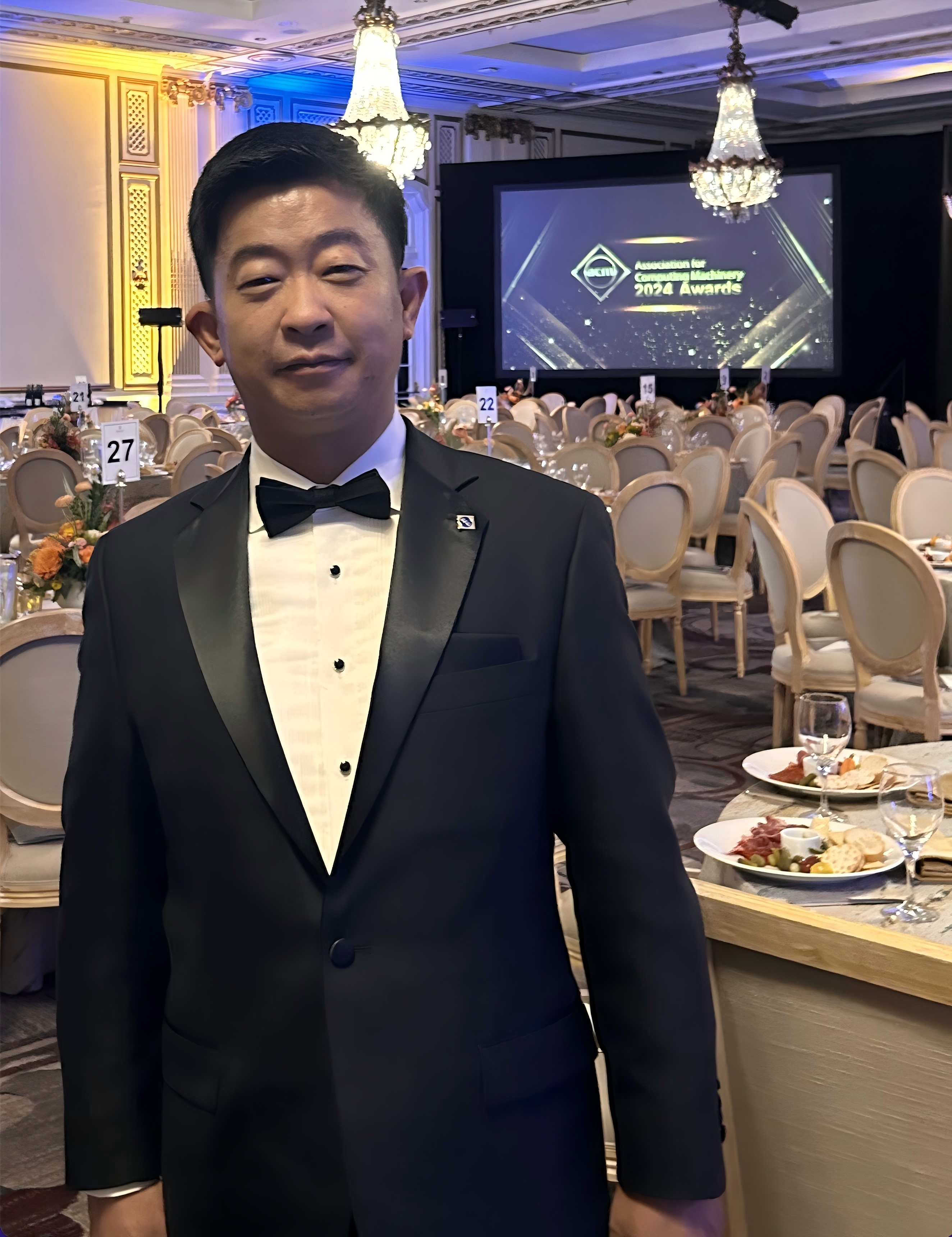}}]{Zhu Han}
	(S’01–M’04-SM’09-F’14) received the B.S. degree in electronic engineering from Tsinghua University, in 1997, and the M.S. and Ph.D. degrees in electrical and computer engineering from the University of Maryland, College Park, in 1999 and 2003, respectively. 
	
	From 2000 to 2002, he was an R\&D Engineer of JDSU, Germantown, Maryland. From 2003 to 2006, he was a Research Associate at the University of Maryland. From 2006 to 2008, he was an assistant professor at Boise State University, Idaho. Currently, he is a John and Rebecca Moores Professor in the Electrical and Computer Engineering Department as well as in the Computer Science Department at the University of Houston, Texas. Dr. Han’s main research targets on the novel game-theory related concepts critical to enabling efficient and distributive use of wireless networks with limited resources. His other research interests include wireless resource allocation and management, wireless communications and networking, quantum computing, data science, smart grid, carbon neutralization, security and privacy. Dr. Han received an NSF Career Award in 2010, the Fred W. Ellersick Prize of the IEEE Communication Society in 2011, the EURASIP Best Paper Award for the Journal on Advances in Signal Processing in 2015, IEEE Leonard G. Abraham Prize in the field of Communications Systems (best paper award in IEEE JSAC) in 2016, IEEE Vehicular Technology Society 2022 Best Land Transportation Paper Award, and several best paper awards in IEEE conferences. Dr. Han was an IEEE Communications Society Distinguished Lecturer from 2015 to 2018 and ACM Distinguished Speaker from 2022 to 2025, AAAS fellow since 2019, and ACM Fellow since 2024. Dr. Han is a 1\% highly cited researcher since 2017 according to Web of Science. Dr. Han is also the winner of the 2021 IEEE Kiyo Tomiyasu Award (an IEEE Field Award), for outstanding early to mid-career contributions to technologies holding the promise of innovative applications, with the following citation: "for contributions to game theory and distributed management of autonomous communication networks."
	
\end{IEEEbiography}
\vspace{-1cm}

\begin{IEEEbiography}[{\includegraphics[width=1in,height=1.25in,clip,keepaspectratio]{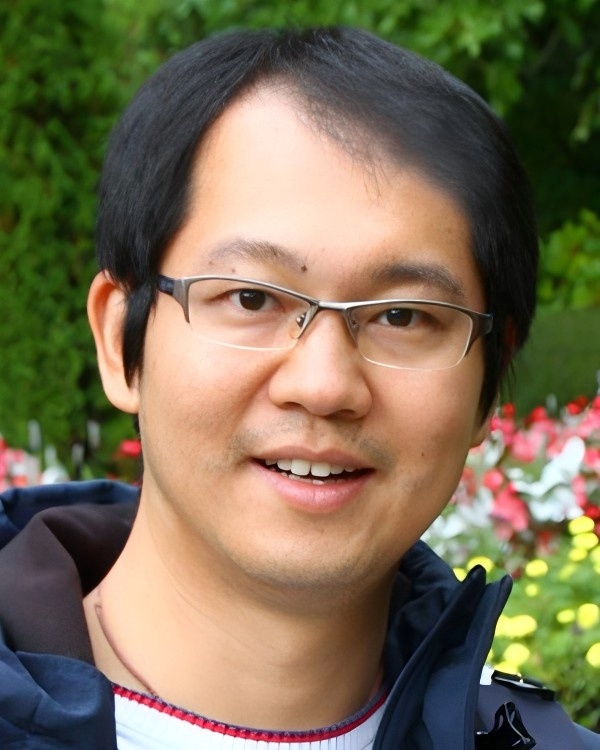}}]{Ning Wang}
	(Member, IEEE) received the B.E. degree in communication engineering from Tianjin University, Tianjin, China, in 2004, the M.A.Sc. degree in electrical engineering from The University of British Columbia, Vancouver, BC, Canada, in 2010, and the	Ph.D. degree in electrical engineering from the University of Victoria, Victoria, BC, Canada, in 2013. 
	
	From 2004 to 2008, he was with the China Information Technology Design and Consulting Institute, as a Mobile Communication System Engineer, specializing in planning and design of commercial mobile communication networks, network traffic analysis, and radio network optimization. From 2013 to 2015, he was a Postdoctoral Research Fellow with the	Department of Electrical and Computer Engineering, The University of British Columbia. Since 2015, he has been with the School of Information Engineering, Zhengzhou University, Zhengzhou, China, where he is currently an Associate Professor. He also holds adjunct appointments with the Department of Electrical and Computer Engineering, McMaster University, Hamilton, ON, Canada and the Department of Electrical and Computer Engineering, University of Victoria. His research interests include resource allocation and security designs of future cellular networks, channel modeling for wireless communications, statistical signal processing, and cooperative wireless communications. He was with the technical program committees of international conferences, including the IEEE GLOBECOM, IEEE ICC, IEEE WCNC, and CyberC. He was the Finalist of the Governor Generals Gold Medal for Outstanding Graduating Doctoral Student with the University of Victoria, Victoria, BC, Canada, in 2013.
\end{IEEEbiography}
\vspace{-1cm}

\begin{IEEEbiography}[{\includegraphics[width=1in,height=1.25in,clip,keepaspectratio]{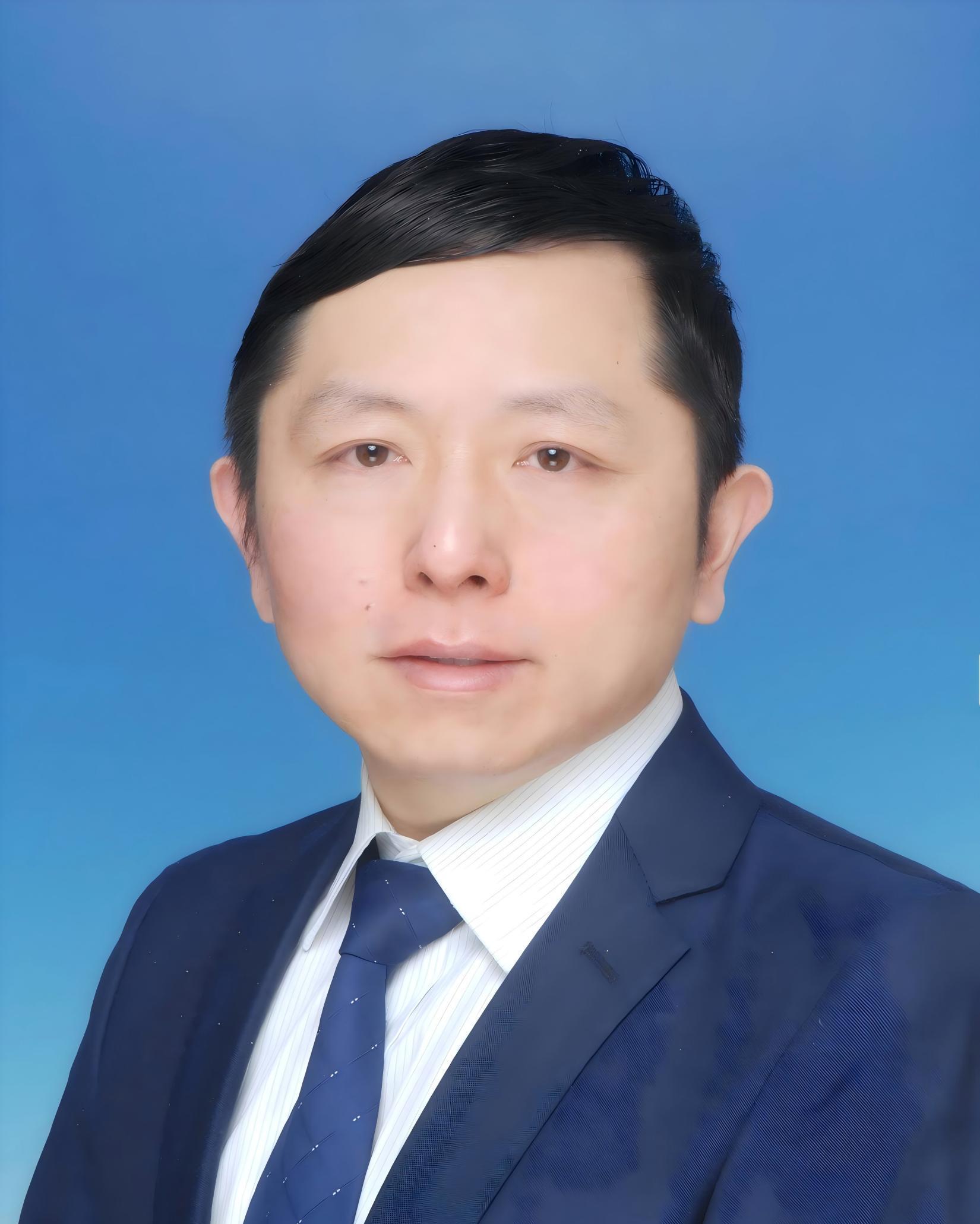}}]{Lei Xiong}
	received the Ph.D. degree from Beijing Jiaotong University, Beijing, China, in 2007. He is currently a Professor and the Vice Deputy Director of the State Key Laboratory of Advanced Rail Autonomous Operation, Beijing Jiaotong University. He has authored/co-authored five books and published over 100 academic research papers in his research area. He holds 15 invention patents. His research interests include dedicated mobile communication system, channel measurement and modeling, and artificial intelligence. Prof. Xiong has been won the First Prize of Science and Technology Award of China Communications Society.
\end{IEEEbiography}

\begin{IEEEbiography}[{\includegraphics[width=1in,height=1.25in,clip,keepaspectratio]{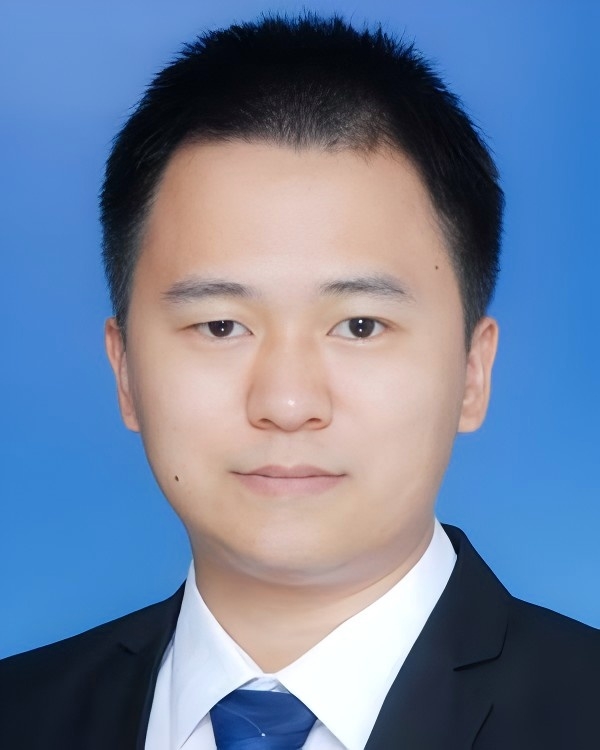}}]{Ruisi He}
	(Senior Member, IEEE) received the B.E. and Ph.D. degrees from Beijing Jiaotong University (BJTU), Beijing, China, in 2009 and 2015, respectively. He is currently a Professor with the State Key Laboratory of Advanced Rail Autonomous Operation and the School of Electronics and Information Engineering, BJTU. He has been a Visiting Scholar with the Georgia Institute of Technology, Atlanta, GA, USA, University of Southern California, Los	Angeles, CA, USA, and Université Catholique de Louvain, Ottignies-Louvain-la-Neuve, Belgium. He	has authored or coauthored eight books, four book chapters, more than 200	journal and conference papers, and several patents. His research interests include wireless propagation channels, railway and vehicular communications, 5G and 6G communications. Dr. He has been the Editor of IEEE Transactions on Communications, IEEE Transactions on Wireless Communications, IEEE Transactions on Antennas and Propagation, IEEE Antennas and Propagation Magazine, IEEE Communications Letters, IEEE Open Journal of Vehicular Technology, and the Lead Guest Editor of the IEEE Journal on Selected Area in Communications and IEEE Transactions	on Antennas and Propagation. He was the Early Career Representative	(ECR) of Commission C, International Union of Radio Science (URSI). He was the recipient of the URSI Issac Koga Gold Medal in 2020, IEEE ComSoc Asia-Pacific Outstanding Young Researcher Award in 2019, URSI Young Scientist Award in 2015, and several best paper awards in IEEE journals and conferences.
\end{IEEEbiography}

\end{document}